\documentclass[aos,MSNbibl,dvips]{arximspdf}
\usepackage{graphicx}
\doi{10.1214/14-AOS1260} % kopijuoti is laisko
\volume{42}
\issue{6}
\pubyear{2014}
\firstpage{2526}
\lastpage{2556}
\docsubty{FLA}

\makeatletter
\newcommand{\rrvert}{\vert}
\newcommand{\rrVert}{\Vert}
\newcommand{\llvert}{\vert}
\newcommand{\llVert}{\Vert}
\renewcommand{\mid}{|}
\newcommand{\PP}{\mathbb{P}}
\newcommand{\EE}{\mathbb{E}}
\newcommand{\R}{\mathbb{R}}
\newcommand{\argmin}{\operatorname{argmin}\limits}
\newcommand{\pa}{\mathrm{pa}}
\newcommand{\do}{\mathrm{do}}
\newtheorem{theo}{Theorem}
\newtheorem{lemm}{Lemma}
\newproclaim{rem}{Remark}
\makeatother

\begin{document}
\begin{frontmatter}

\title{CAM: Causal additive models, high-dimensional order search and
penalized regression}
\runtitle{CAM: Causal additive models}

\begin{aug}
\author[A]{\fnms{Peter}~\snm{B\"uhlmann}\ead[label=e1]{buhlmann@stat.math.ethz.ch}},
\author[A]{\fnms{Jonas} \snm{Peters}\thanksref{T2}\corref{}\ead[label=e2]{peters@stat.math.ethz.ch}}
\and
\author[A]{\fnms{Jan} \snm{Ernest}\thanksref{T3}\ead[label=e3]{ernest@stat.math.ethz.ch}\ead[label=u1,url]{http://stat.ethz.ch}}
\runauthor{P. B\"uhlmann, J. Peters and J. Ernest}
\affiliation{ETH Z\"urich}
\address[A]{Seminar for Statistics\\
ETH Z\"urich\\
R\"amistrasse 101\\
8092 Z\"urich\\
Switzerland\\
\printead{e1}\\
\phantom{E-mail: }\printead*{e2}\\
\phantom{E-mail: }\printead*{e3}\\
\printead{u1}}
\end{aug}
\thankstext{T2}{Supported by the People Programme (Marie Curie
Actions) of the European Union's Seventh Framework Programme
(FP7/2007-2013) under REA Grant agreement no. 326496.}
\thankstext{T3}{Supported in part by the Swiss National Science
Foundation Grant no. 20PA20E-134493.}

% HISTORY:
\received{\smonth{10} \syear{2013}}
\revised{\smonth{7} \syear{2014}}

% ABSTRACT
%
\begin{abstract}
We develop estimation for potentially high-dimensional additive
structural equation models.
A key component of our approach is to decouple order search among the
variables from feature or
edge selection in a directed acyclic graph encoding the causal
structure. We show that the former
can be done with nonregularized (restricted) maximum likelihood
estimation while the latter can be
efficiently addressed using sparse regression techniques. Thus, we
substantially simplify the problem
of structure search and estimation for an important class of causal
models. We establish consistency
of the \mbox{(restricted)} maximum likelihood estimator for low- and
high-dimensional scenarios, and we also allow
for misspecification of the error distribution. Furthermore, we develop
an efficient computational
algorithm which can deal with many variables, and the new method's
accuracy and performance is illustrated on simulated and real data.
\end{abstract}

% KEYWORDS
% Pirmas kwd is didziosios raides
%
\begin{keyword}[class=AMS]
\kwd[Primary ]{62G99}
\kwd{62H99}
\kwd[; secondary ]{68T99}
\end{keyword}
\begin{keyword}
\kwd{Graphical modeling}
\kwd{intervention calculus}
\kwd{nonparametric regression}
\kwd{regularized estimation}
\kwd{sparsity}
\kwd{structural equation model}
\end{keyword}
\end{frontmatter}

\setcounter{footnote}{2}

%s1 #&#
\section{Introduction}

Inferring causal relations and effects is an ambitious but important task
in virtually all areas of science. In absence of prior information about
underlying structure, the problem is plagued, among other things, by
identifiability issues \cite{pearl00,sgs00}, cf., and the sheer size of the
space of possible models, growing super-exponentially in the number
of variables, leading to major challenges with respect to computation and
statistical accuracy.
%We will address here the latter \Jonas{'the latter' unclear?} issue:
Our approach is generic, taking advantage of the tools in
sparse regression techniques \cite{hastetal09,pbvdg11}, cf., which have been
successively established in recent years.

More precisely, we consider $p$ random variables $X_1,\ldots,X_p$ whose
distribution is Markov with respect to an underlying causal directed
acyclic graph (causal DAG). We assume that all variables are observed,
that is, there are no
hidden variables, and that the causal influence diagram does not allow for
directed cycles. Generalizations to include hidden variables, for example,
unobserved confounders, or directed
cycles are briefly discussed in Section~\ref{subsecextensions}. To
formalize a
model, one can
use the concepts of graphical modeling \cite{lauritzen96}, cf., or
structural equation models~\cite{pearl00}, cf. The approaches are
equivalent in the nonparametric or multivariate Gaussian case, but
this is
not true anymore when placing additional restrictions which can be very
useful \cite{shim06,petersetal13,petbu13}. We use here the framework
of structural
equation models.

%s1.1 #&#
\subsection{Problem and main idea}

Our goal is estimation and structure learning for structural
equation models, or of the corresponding Markov equivalence class of an
underlying DAG. In particular, we focus on causal additive models, that is,
the structural equations are additive in the variables and error terms. The
model has the nice property that the underlying structure and the
corresponding parameters are identifiable from the observational
distribution. Furthermore, we can view it as an extension of linear
Gaussian structural equation models by allowing for nonlinear additive
functions.

In general, the problem of structure learning (and estimation of
corresponding parameters) can be addressed by a variety of
algorithms and methods: in the frequentist setting, the most widely used
procedures for structure learning (and corresponding parameters) are greedy
equivalence search for computing the \mbox{BIC-}regularized maximum likelihood
estimator \cite{chick02} or the PC-algorithm using multiple conditional
independence testing \cite{sgs00}. However, for the latter, the constraint
of additive structural equations cannot be (easily) respected, and
regarding the former, maximum
likelihood estimation among all (e.g., linear Gaussian) DAG models is
computationally
challenging and statistical guarantees for high-dimensional cases (and for
uniform convergence with respect to a class of distributions)
are only available under rather strong assumptions \cite{sarpet12}.

Our proposed approach for estimation and selection of additive structural
equation models is based on the following simple idea which is
briefly mentioned and discussed in \cite{teykol05} and
\cite{schmidt07}. If the order among the variables would be known, the
problem boils down to
variable selection in multivariate (potentially nonlinear) regression; see
formula (\ref{triangular}). The latter is very well understood: for
example, we can follow the route of hypothesis testing in additive models,
or sparse regression can be used
%as it
%belongs to the area of sparse regression \Jonas{Mhh. Just to make
%sure: In the low-dim setting, we do not need high-dim selection
%techniques, do we? In the high-dim setting, we are starting with the
%prel. neighborhood sel. But then, we again do not necessarily need
%high-dim techniques for the variable selection (pruning). But, we can
%use the (group) lasso nevertheless, right?}: for example, we can use
%the Lasso
for
%linear regression
additive models \cite{YI06,ravik09,meieretal09}. Thus, the only remaining
task is to
estimate the order among the variables. We show here that this
can be done via the maximum likelihood principle, and we establish its
consistency. In particular, for low
or ``mid''-dimensional problems, there is no need to consider a penalized
likelihood approach. The same holds true for high-dimensional settings when
using a preliminary neighborhood selection and then employing a
corresponding restricted maximum likelihood estimator. Therefore, we can
entirely decouple the issue of order estimation without regularization and
variable selection in sparse regression with appropriate
regularization. This makes our approach very
generic, at least within the framework where the underlying DAG and a
corresponding order of the variables are identifiable
from the joint distribution. Empirical results in Section~\ref{secnumerical}
support that we
can do much more accurate estimation than for nonidentifiable models such
as the popular linear Gaussian structural equation model.
%from a fundamental (but yet not practical) computational point of
%view, we gain in terms
%of computational complexity $\mathcal{ C}(p)$: $\log({\mathcal C}(p)) =
%O(p^2)$
%when searching over all DAG structures versus $\log({\mathcal C}(p)) =
%O(p
On the superficial level, our approach can be summarized as follows:
\begin{longlist}[3.]
\item[1.] Mainly for high-dimensional settings: preliminary neighborhood
selection for estimating a superset of the skeleton of the underlying
DAG. This is done by additive regression of one variable against all
others. See Section~\ref{secrestrMLE}.
\item[2.] Order search for the variables (or best permutation for the indices
of the variables) using (restricted) maximum likelihood estimation
based on an
additive structural equation model with Gaussian errors: the restricted
version is employed if the preliminary neighborhood selection in step~1 is
used, and the order search is then restricted to the structure of the superset
of the skeleton. See Sections~\ref{subsecmle} and~\ref{secrestrmle}.
\item[3.] Based on the estimated order of the variables in step~2, sparse
additive regression is used for estimating the functions in an additive
structural equation model. See Section~\ref{subsecvarsel}.
\end{longlist}

%s1.2 #&#
\subsection{Related work}

We consider (nonlinear) additive structural equation models. As natural
extensions of linear structural equation models, they are attractive
for many
applications; see Imoto, Goto and Miyano \cite{imoto02}. Identifiability results for this model class have been
recently derived \cite{moo09,petersetal13}. The approach in \cite
{moo09} is
based on conditional independence testing and is limited to small
dimensions with a few variables only. Instead of multiple testing of
conditional independencies, we propose and develop maximum likelihood
estimation in a semiparametric additive structural equation model with
Gaussian noise variables: fitting such a model is often appropriate in
situations where the sample size is not too large, and we present here for
the first time the practical feasibility of fitting additive models in the
presence of
many variables. An extension of our additive structural equation
model with Gaussian errors to the case with a nonparametric specification
of the error distribution is presented in \cite{nowzopb13}, but the
corresponding maximum likelihood estimator is analyzed (and feasible) for
problems with a small number of variables only. When the order of the
variables is known, which is a much simpler and different problem than what
we consider here, \cite{vooretal13} provide consistency results for
additive structural equation models.

A key aspect of our method is that we decouple regularization for feature
selection and order estimation with nonregularized (restricted) maximum
likelihood. The former is a well understood subject thanks to the broad
literature in sparse regression and related techniques
\cite{tibs96,mebu06,YI06,zhaoyu06,zou06,wain09}, cf. Regarding the latter
issue about order selection, a recent analysis in \cite{vdg13} extends our
low-dimensional consistency result for the (nonrestricted) maximum
likelihood estimator to the scenario where the
number of variables can grow with sample size, in the best case essentially
as fast as $p = p(n) = o(n)$. The treatment of the high-dimensional case
with a restricted maximum likelihood approach is new here, and we also
present the first algorithm and empirical results for fitting low-
and high-dimensional causal additive models (CAMs).

All proofs are provided in the supplemental article \cite{CAMsuppl14}.

%s2 #&#
\section{Additive structural equation models} \label{secmle}

Consider the general structural equation model (SEM):
\[
X_j = f_{j}(X_{\pa_D(j)},\varepsilon_j),
\qquad\varepsilon_1,\ldots,\varepsilon_p
\mbox{ (mutually) independent},
\]
where $\pa_D(j)$ denotes the set of parents for node $j$ in DAG
$D$ and $f_j$ is a function from $\R^{\llvert \pa_D(j)\rrvert +1} \to
\R$. Thus, a
SEM is
specified by an underlying (causal) structure in terms
of a DAG $D$, the functions $f_j(\cdot)$ $(j=1,\ldots,p)$ and the
distributions of $\varepsilon_j$ $(j=1,\ldots,p)$. Most parts of this
paper can be
interpreted in absence of causal inference issues: clearly though, the main
motivations are understanding models and developing novel procedures
allowing for causal or interventional
statements, and if we do so, we always assume that the structural equations
remain unchanged under interventions at one or several variables
\cite{pearl00}, cf. The model above is often too general,
due to problems of identifiability and the difficulty of estimation (curse
of dimensionality) of functions in several variables.

Our main focus is on a special (and more practical) case of the model
above, namely the additive SEM with potentially misspecified Gaussian errors:
%e1 #&#
%
\begin{eqnarray}\label{SEMadd}
&& X_j = \sum_{k \in\pa_D(j)}
f_{j,k} (X_k) + \varepsilon_j,
\nonumber
\\[6pt]
\\[-20pt]
\eqntext{\displaystyle \varepsilon_1,\ldots,\varepsilon_p \mbox{ independent with } \varepsilon_j \sim{\mathcal N}\bigl(0,\sigma_j^2
\bigr), \sigma_j^2 >0\ (j=1,\ldots,p),}
\\
\eqntext{\displaystyle \EE\bigl[f_{j,k}(X_k)\bigr] = 0\mbox{ for all } j,k,\hspace*{172pt}}
\end{eqnarray}
where $f_{j,k}(\cdot)$ are smooth functions from $\R\to\R$.
A special
case thereof is the linear Gaussian SEM
%e2 #&#
%
\begin{eqnarray}\label{SEMlinGauss} && X_j = \sum_{k \in\pa_D(j)}
\beta_{j,k} X_k + \varepsilon_j,
\nonumber\\[-8pt]\\[-8pt]
\eqntext{\varepsilon_1,\ldots,\varepsilon_p \mbox{ independent with } \varepsilon_j \sim{\mathcal N}\bigl(0,\sigma_j^2
\bigr), \sigma_j^2 >0\ (j=1,\ldots,p).}
\end{eqnarray}
Although model (\ref{SEMlinGauss}) is a special case of (\ref{SEMadd}),
there are interesting differences with respect to identifiability. If all
functions $f_{j,k}(\cdot)$ are nonlinear, the DAG is identifiable from
the distribution $P$ of $X_1,\ldots,X_p$ \cite{petersetal13}, Corollary~31. We explicitly state this result as a lemma since we
will make use of it later on.

%le1 #&#
%
\begin{lemm}[(Corollary~31 in \cite{petersetal13}\footnote
{Corollary~31 in \cite{petersetal13} contains a slightly different
statement using ``nonlinear'' instead of ``nonconstant''. The proof,
however, stays exactly the same.})]\label{lemid}
Consider a distribution $P$ that is generated by model~(\ref{SEMadd}) with
DAG $D$ and nonlinear, three times differentiable functions $f_{j,k}$. Then
any distribution $Q$ that is generated by~(\ref{SEMadd}) with a different
DAG $D' \neq D$ and nonconstant, three times differentiable functions
$f_{j,k}'$ is different from $P$: we have $Q \neq P$.
\end{lemm}
This result does not hold, however, for a general SEM or for a linear
Gaussian SEM as in (\ref{SEMlinGauss}); one can then only identify
the Markov equivalence class of the DAG $D^0$, assuming faithfulness.
An exception arises when assuming same error variances $\sigma_j^2
\equiv
\sigma^2$ for all $j$ in (\ref{SEMlinGauss}) which again implies
identifiability of the DAG $D^0$ from $P$ \cite{petbu13}. In the
sequel, we
consider the fully identifiable case of model (\ref{SEMadd}).
%and specialize to \eqref{SEMlinGauss} at some places.

%s2.1 #&#
\subsection{The likelihood}

We slightly re-write model (\ref{SEMadd}) as
%e3 #&#
%
\begin{eqnarray}
\label{modsem} && X_j = \sum_{k \in\pa_D(j)}
f_{j,k}(X_k) + \varepsilon_j = \sum
_{k\neq j} f_{j,k}(X_k) +
\varepsilon_j\qquad (j=1,\ldots,p),
\nonumber
\\
\eqntext{\displaystyle f_{j,k}(\cdot) \not\equiv0 \mbox{ if and only if there is a directed edge }k \to j\mbox{ in }D,}
\\[-8pt]\\[-8pt]
\eqntext{\displaystyle \EE\bigl[f_{j,k}(X_k)\bigr] = 0\mbox{ for all } j,k,\hspace*{144pt}}
\\
\eqntext{\displaystyle \varepsilon_1,\ldots,\varepsilon_p\mbox{ independent and } \varepsilon_j \sim{\mathcal N}\bigl(0,\sigma_j^2
\bigr), \sigma_j^2 >0.\hspace*{43pt}}
\end{eqnarray}
Note that the structure of the model, or the so-called active set,
$\{(j,k); f_{j,k} \not\equiv0\}$ is identifiable from the distribution
$P$ \cite{petersetal13}, Corollary~31.
Denote by $\theta$ the infinite-dimensional parameter with additive
functions and error variances, that is,
\[
\theta= (f_{1,2},\ldots, f_{1,p},f_{2,1},\ldots,f_{p,p-1},\sigma_1,\ldots,\sigma_p).
\]
Furthermore, we denote by $D^0$ the true DAG and by $\theta^0$ (and $\{
f_{j,k}^0\}$, $\{\sigma_j^0\}$) the
true infinite-dimensional parameter(s) corresponding to the data-ge\-nerating
true distribution. We use this notation whenever it is appropriate to make
statements about the true underlying DAG or parameter(s).

The density $p_{\theta}(\cdot)$ for the
model (\ref{modsem}) is of the form
\[
\log\bigl(p_{\theta}(x)\bigr) = \sum_{j=1}^p
\log\biggl(\frac{1}{\sigma_j} \varphi\biggl(\frac{x_j - \sum_{k\neq
j} f_{j,k}(x_k)}{\sigma
_j} \biggr) \biggr),
\]
where $\varphi(\cdot)$ is the density of a standard normal distribution.
Furthermore,
\[
\sigma_j^2 = \EE\biggl[\biggl(X_j - \sum
_{k\neq j} f_{j,k}(X_k)
\biggr)^2\biggr],
\]
and the expected negative log-likelihood is
\[
\EE_{\theta}\bigl[-\log p_{\theta}(X)\bigr] = \sum
_{j=1}^p \log(\sigma_j) + C,\qquad C = p
\log(2 \pi)^{1/2} + p/2.
\]

%s2.2 #&#
\subsection{The function class}
We assume that the functions in model (\ref{SEMadd}) or (\ref{modsem})
are from a class of smooth functions:
$\mathcal{F}$ is a subset of $L_2(P_j)$, where $P_j$ is the marginal distribution for
any $j=1, \ldots, p$; assume  that it is closed with respect to the $L_2(P_j)$ norm. Furthermore,
\[
\mathcal{F} \subseteq \bigl\{f\dvtx \R \to \R, f \in C^\alpha, \EE\bigl[f(X)\bigr] = 0\bigr\},
\]
where $C^{\alpha}$ denotes the space of $\alpha$-times differentiable
functions and the random variable $X$ is a placeholder for the
variables $X_j$, $j=1,\ldots ,p$. Note that this is a slight abuse of
notation since $\mathcal{F}$ does not specify the variable $X$; it becomes clear from the context.
%
%where $C^{\alpha}$ denotes the space of $\alpha$-times differentiable
%functions.
% implicitly when assuming entropy with bracketing since we need to make
% sure that the brackets exist in $L_2(P)$.}

Consider also basis functions $\{b_r(\cdot); r=1,\ldots,a_n\}$ with $a_n
\to\infty$ sufficiently slowly, for example,
B-splines or regression splines. Consider further the space
%(closed with respect to the $L_2(P_i)$ norm, where $P_i$ is the
%marginal distribution for any $i=1, \ldots, p$)
%what do you think?}
%e4 #&#
%
\begin{equation}\label{basis-exp}
\mathcal{F}_n = \Biggl\{f \in \mathcal{F}, f = c + \sum_{r=1}^{a_n} \alpha_r b_r(\cdot)
\mbox{ with } c, \alpha_r \in \R (r=1,\ldots, a_n)\Biggr\}.
\end{equation}
We allow for constants $c$ to enforce mean zero for the whole function. Furthermore,
the basis functions can be the same for all variables $X_j$, $j=1,\ldots ,p$.

For theoretical analysis, we assume that ${\mathcal F}_n$ is
deterministic and
does not depend on the data. Then, ${\mathcal F}_n$ is closed. %
% sagen wir, dass ``in practice'' koennen die functions von den
%Variablen
% und der order abhaengen}
Furthermore, the space of additive functions is denoted by
\begin{eqnarray*}
{\mathcal F}^{\oplus\ell} &=& \Biggl\{f\dvtx  \R^\ell\to\R; f(x) = \sum
_{k=1}^\ell f_k(x_k),
f_k \in{\mathcal F}\Biggr\},
\\
{\mathcal F}_n^{\oplus\ell} &=& \Biggl\{f\dvtx  \R^\ell\to
\R; f(x) = \sum_{k=1}^\ell
f_k(x_k), f_k \in{\mathcal
F}_n\Biggr\},
\end{eqnarray*}
where $\ell=2,\ldots,p$. Clearly, ${\mathcal F}_n^{\oplus\ell}
\subseteq{\mathcal
F}^{\oplus\ell}$.
For $f \in{\mathcal F}^{\oplus\ell}$ we denote by $f_k$ its $k$th
additive function.

%Later, we consider projections of distributions onto the spaces
%${\mathcal F}^{\oplus\ell}$ and ${\mathcal F}_n^{\oplus\ell}$;
%see~(\ref{eqproje}).
%The following lemma guarantees that both of these spaces are closed
%with respect to the $L_2$ norm.
%Intuitively, this result implies that the distribution generated
%by~(\ref{SEMadd}) cannot be approximated arbitrarily well with a
%wrong DAG; see Lemma~\ref{lemid}. As with other theoretical results,
%Lemma~\ref{lemclosed} is proved in the supplemental article \cite
%{CAMsuppl14}.
%%le2 #&#
%%
%Let the distribution $P$ be generated according to~(\ref{SEMadd}) and
%assume that there is a density $p$ with respect to the Lebesgue
%measure. %\Peter{Density with respect to which measure?}
%Assume further that the mean square contingency \cite{Renyi1959}, for example, is finite, that is,
%%
%%
%for any pair $j \neq k$. For any subset $I \subseteq\{1, \ldots, p\}$
%of $\ell$ variables the spaces ${\mathcal F}^{\oplus\ell}$ and
%${\mathcal F}_n^{\oplus\ell}$ are then closed with respect to the
%$L_2(P_{I})$ norm. Here, $P_I$ denotes the marginal distribution over
%all variables in $I$.
%%
%The question of closedness of additive models has also been studied in

In our definitions, we assume that the functions in $\mathcal{F}$ and $\mathcal{F}_n$ have expectation zero.
Of course, this depends on the variables in the arguments of the functions. For example, when
requiring $\EE[f(X_j)] = 0$ for $f \in \mathcal{ F}$, the function class
$\mathcal{ F} = \mathcal{ F}_j$ depends on the index $j$ due to the mean zero requirement; and
likewise $\mathcal{ F}^{\oplus \ell}$ depends on the indices of the variables occurring in the
corresponding additive function terms. We drop this additional dependence on the index of variables as it does not cause any problems in methodology or theory.

Later, we consider projections of distributions onto the spaces $\mathcal{ F}^{\oplus \ell}$ and $\mathcal{ F}_n^{\oplus \ell}$, see~(\ref{eqproje}).
We assume throughout the paper that these spaces are closed with respect to the $L_2$ norm. The following
Lemma~\ref{lemclosed} guarantees this condition by requiring an analogue of a minimal eigenvalue assumption.

\begin{lemm} \label{lemclosed}
Let the distribution $P$ be generated according to~(\ref{SEMadd}) and
assume that there is a $\phi^2 > 0$ such that for all $\gamma \in \R^p$
\[
\Biggl\| \sum_{j = 1}^p \gamma_j f_j(X_j) \Biggr\|^2_{L_2} \geq \phi^2
\|\gamma\|^2\qquad\mbox{for all } f_j \in \mathcal{F} %\text{ }(\text{or } \mathcal{F}_n)
\mbox{ with } \bigl\|f_j(X_j)\bigr\|_{L_2} = 1.
\]
For any subset $I \subseteq \{1, \ldots, p\}$ of $\ell$ variables the spaces $\mathcal{ F}^{\oplus \ell}$
and $\mathcal{ F}_n^{\oplus \ell}$ are then closed with respect to the $L_2(P_{I})$ norm. Here, $P_I$ denotes the marginal distribution over all variables in $I$.
\end{lemm}

The question of closedness of additive models has also been studied in \cite{breifried85}, for example; see also \cite{Renyi1959}.

%s2.3 #&#
\subsection{Order of variables and the likelihood}

We can permute the variables, inducing a different ordering; in the sequel,
we use both terminologies, permutations and order search, which mean the
same thing. For a
permutation $\pi$ on $\{1,\ldots,p\}$,
define
\[
X^{\pi},\qquad X^{\pi}_j = X_{\pi(j)}.
\]
%
%formula
% ``is of the form:'' below.}\Peter{Verstehe ich nicht: meinst Du $x^{
There is a canonical correspondence between permutations and fully
connected DAGs: for any permutation $\pi$, we can construct a DAG
$D^{\pi}$,
in which each variable $\pi(k)$ has a directed arrow to all $\pi(j)$ with
$j>k$. The node $\pi(1)$ has no parents and is called the source node. For
a given DAG $D^0$, we define the set of true permutations as
\[
\Pi^0 = \bigl\{\pi^0; \mbox{ the fully connected DAG }
D^{\pi^0} \mbox{ is a super-DAG of } D^0\bigr\},
\]
where a super-DAG of $D^0$ is a DAG whose set of directed edges is a
superset of the one corresponding to $D^0$. %\Peter{Habe das ergaenzt:
%OK?}
If the true DAG $D^0$ is not fully connected, there is typically more than
one true order or permutation, that is the true order is typically not unique.
It is apparent that any true ordering or permutation $\pi^0$ allows
for a lower-triangular
(or autoregressive) representation of the model in (\ref{modsem}):
%e5 #&#
%
\begin{equation}
\label{triangular} X^{\pi^0}_j = \sum
_{k=1}^{j-1} f^{\pi^0}_{j,k}
\bigl(X^{\pi^0}_k\bigr) + \varepsilon^{\pi^0}_j\qquad
(j=1,\ldots,p),
\end{equation}
where $f^{\pi^0}_{j,k}(\cdot) = f^0_{\pi^0(j),\pi^0(k)}(\cdot)$ and
$\varepsilon^{\pi^0}_j = \varepsilon^0_{\pi^0(j)}$, that is, with
permuted indices in terms of
the original quantities in (\ref{modsem}). %Note that $\pi(1)$ has no
%parents, called the source node.
%have any parents? As apposed to $\pi^{-1}(1)$ being the source node.}
%source node?}
%: it is an element of
If all functions $f_{j,k}(\cdot)$ are nonlinear, the
set of true permutations is identifiable from the distribution \cite{petersetal13}, Corollary~33, and
$\Pi^0$ consists
of all orderings of the variables which allow for a lower-triangular
representation~(\ref{triangular}). %are
%consistent with the arrow directions in the true DAG $D^0$.
We will exploit this fact in order to provide a consistent estimator
$\hat\pi_n$ of the ordering: under suitable assumptions the
probability that $\hat\pi_n \in\Pi^0$ converges to one.

%re1 #&#
%
\begin{rem} \label{rem1}
For the linear Gaussian SEM (\ref{SEMlinGauss}), all orderings allow
for a
lower-triangular representation~(\ref{triangular}), even those that
are not
in $\Pi^0$.
%consists of all permutations
%and corresponding orderings of the variables which are
%consistent with the arrow directions in a DAG of the Markov equivalence
%class of the true DAG $D^0$.
%since we can represent every Gaussian distribution as in
Thus, we cannot construct a consistent estimator in the above sense.
However, assuming faithfulness of the true
distribution, the orderings of
variables which are consistent with the arrow directions in a DAG of the
Markov equivalence
class of the true DAG $D^0$ lead to sparsest representations with fewest
number of nonzero coefficients.% (Any sparser representation contains
%a pair of nodes with zero coefficients that are connected in the
%Markov equivalence class of $D^0$; this contradicts faithfulness.)
%See for example \cite{sarpet12}.
%zitierten paper ist das auch nicht so genau drin, glaube ich.}

In principle, one can check whether the data come from a linear Gaussian
SEM. Lemma~\ref{lemid} guarantees that if this is case, there is no
CAM with nonlinear functions yielding the same distribution.
%..., there is no nonlinear CAM yielding the same distribution\\
%or\\
%..., there is no nonlinear additive noise SEM yielding the same
%distribution?
%}
Thus, if the structural equations of the estimated DAG look linear
with Gaussian noise, one could
decide to output the
Markov equivalence class instead of the DAG. One would need to quantify
closeness to linearity and Gaussianity with, for example, a test: this
would be
important for practical applications, but its precise implementation lies
beyond the scope of this work.
\end{rem}

In the\vspace*{1pt} sequel, it is helpful to consider the true underlying parameter
$\theta^0$ with corresponding nonlinear function $f_{j,k}^0$ and error
variances $(\sigma_j^0)^2$. For any permutation $\pi\notin\Pi^0$, we
consider
%a
%projected model representation (in fact, this representation is not
%needed
%to look at!) in the space of additive SEMs with Gaussian errors:
%& &Y^{\pi}_j = \sum_{k=1}^{j-1} f^{\pi}_{j,k}(Y^{\pi}_k) + \eta^{\pi}_j
% %(j=1,\ldots,p),\\
%& &\eta^{\pi}_1,\ldots,\eta^{\pi}_p \mbox{independent and} \eta^{
%{\mathcal N}(0,(\sigma^{\pi}_j)^2).
the projected parameters, defined as
\[
\theta^{\pi,0} = \argmin_{\theta^{\pi}} \EE_{\theta^0}\bigl[- \log
\bigl(p_{\theta^{\pi}}^{\pi}(X)\bigr)\bigr],
\]
where the density
$p^{\pi}_{\theta^{\pi}}$ is of the form
\[
\log\bigl(p^{\pi}_{\theta^{\pi}}(x)\bigr) = \log\bigl(p_{\theta^{\pi}}
\bigl(x^{\pi
}\bigr)\bigr) = \sum_{j=1}^p
\log\biggl(\frac{1}{\sigma^{\pi}_j} \varphi\biggl(\frac{x^{\pi}_j -
\sum_{k=1}^{j-1}
f^{\pi}_{j,k}(x^{\pi}_k)}{\sigma^{\pi}_j} \biggr) \biggr).
\]
%
%Note that we could alternatively (and less obvious to understand)
%write:
%p_{\theta^{\pi}}^{\pi}(x) = p_{\theta^{\pi}\circ\pi^{-1}}(x^{\pi}),
%where the $j$th component is $(\theta^{\pi}\circ\pi^{-1})_j =
%(\theta^{\pi})_{\pi^{-1}(j)}$ (because of our notation where
%$p_{\theta}(\cdot)$ is associated with an ordering of $\theta$ as
%well).
%
%The expressopm above is simply the (negative logarithm of) the density
%when
%could construct the best approximating backwards model as:
%& & \eps^{\pi}_1,\ldots,\eps^{\pi}_p \mbox{independent and} %\eps^{
%{\mathcal N}(0,(\sigma^{\pi}_j)^2),\\
%& &X^{\pi}_1 = \eps^{\pi}_1,\\
%& &X^{\pi}_2 = f^{\pi}_{2,1}(X^{\pi}_1) + \eps^{\pi}_2,\\
%& &X^{\pi}_j = \sum_{k=1}^{j-1} f^{\pi}_{j,k}(X^{\pi}_k) + \eps^{
%& &X^{\pi}_p = \sum_{k=1}^{p-1} f^{\pi}_{j,k}(X^{\pi}_k) + \eps^{
%
%We write
%-\log(p^{\pi}_{\theta^{\pi}}(\cdot)),
%and
%P\rho_{\theta} = \EE_P[\rho_{\theta}(X)], P\rho_{\pi,\theta^{\pi}} =
%We always consider $P$ as the true distribution with density
%$p_{\theta_0}$. Note that $P \rho_{\theta^0} = P
(Note that
%the definition of $\theta^{\pi,0}$ can be used for any
%permutation $\pi$:
if $\pi\in\Pi^0$, then $\theta^{\pi,0} = \theta^0$.)
% correct? or should it be sth like $\theta^{\pi,0} = \theta_{\pi}^0$ -
%if
% the latter is defined}
%$\theta^{\pi,0}$ nie bis auf die Koordinaten genau definiert haben:
%OK?}
For such a misspecified model with wrong order $\pi\notin\Pi^0$, we
have %\Jan{Wir haben alle ar...} \Jonas{...gmins geaendert. Ok?}
%e6 #&#
%
\begin{eqnarray}
\label{eqproje} \bigl\{f_{j,k}^{\pi,0}\bigr\}_{k=1,\ldots, j-1} &=&
\argmin_{g_{j,k} \in
\mathcal{F}, k =1, \ldots, j-1} \EE_{\theta^0}\Biggl[\Biggl(X^{\pi}_j
- \sum_{k=1}^{j-1} g_{j,k}
\bigl(X^{\pi}_k\bigr)\Biggr)^2\Biggr]
\nonumber\\[-8pt]\\[-8pt]\nonumber
&=& \argmin_{g_j \in\mathcal{F}^{\oplus j-1}} \EE_{\theta
^0}\bigl[\bigl(X^{\pi}_j
- g_j\bigl(X^{\pi}_1, \ldots,X^{\pi}_{j-1}\bigr)\bigr)^2\bigr].
\end{eqnarray}
%
%We do not enforce that $\EE[f_{j,k}^{\pi,0}(X_k^{\pi})] = 0$ since the
%minimization below does not depend whether we shift constants between
%different functions $f_{j,k}^{\pi,0}(\cdot)$ for different indices
%$k$.
It holds that
%e7 #&#
%
\begin{eqnarray}
\label{add1} \bigl(\sigma_j^{\pi,0}\bigr)^2 &=&
\argmin_{\sigma^2} \Biggl(\log(\sigma) + \frac{1}{2 \sigma^2} \EE
_{\theta^0}
\Biggl[\Biggl(X^{\pi}_j - \sum_{k=1}^{j-1}
f_{j,k}^{\pi,0}\bigl(X^{\pi}_k\bigr)
\Biggr)^2\Biggr] \Biggr)
\nonumber\\[-8pt]\\[-8pt]\nonumber
&=& \EE_{\theta^0}\Biggl[\Biggl(X^{\pi}_j - \sum
_{k=1}^{j-1} f_{j,k}^{\pi,0}
\bigl(X^{\pi}_k\bigr)\Biggr)^2\Biggr].
\end{eqnarray}
The two displayed formulae above show that autoregression with the
wrong order
$\pi$ leads to the projected parameters $\{f_{j,k}^{\pi,0}\}$ and
$\{(\sigma_j^{\pi,0})^2\}$. Finally, we obtain
\[
\EE_{\theta^0}\bigl[-\log\bigl(p^{\pi}_{\theta^{\pi,0}}(X)\bigr)
\bigr] = \sum_{j=1}^p \log\bigl(
\sigma_{j}^{\pi,0}\bigr) + C,\qquad C = p \log(2 \pi)^{1/2} +
p/2.
\]
All true permutations $\pi\in\Pi^0$ correspond to super DAGs of the true
DAG and, therefore, all of them lead to the minimal expected
log-likelihood\break
$\EE_{\theta^0}[-\log(p^{\pi}_{\theta^{\pi,0}}(X))] =
\EE_{\theta^0}[-\log(p_{\theta^{0}}(X))]$.
The permutations $\pi\notin\Pi^0$, however, cannot lead to a smaller
expected negative log-likelihood
(since it would lead to a negative %smaller
KL-divergence between the true
and best projected distribution).
%((larger KL divergence between the true and best projected
%distribution)
Let us therefore define
%we want to use this as a definition, we have to change other
%equations, too.}
%e8 #&#
%
\begin{equation}
\label{identifiability} \xi_p:= \min_{\pi\notin\Pi^0} p^{-1}
\bigl(\EE_{\theta^0}\bigl[-\log\bigl(p^{\pi}_{\theta^{\pi,0}}(X)\bigr)
\bigr] - \EE_{\theta^0}\bigl[-\log\bigl(p_{\theta^0}(X)\bigr)\bigr]\bigr)
\geq0.
\end{equation}
%
% old
%(\EE_{\theta^0}[-\log(p^{\pi}_{\theta^{\pi,0}}(X))] -
%Linear Gaussian SEMs do not allow for identifiability of the ordering
%and thus $\xi_p = 0$, see Remark~\ref{rem1}.
If all true functions $f_{j,k}^0$ are nonlinear, we obtain $\xi_p >
0$ as
follows.
%a direct consequence of the identifiability stated in Lemma~
%and the closedness of the function spaces $\mathcal{F}^{\oplus j}$, see
%Lemma~\ref{lemclosed}.

%le3 #&#
%
\begin{lemm} \label{lemxip}
Consider a distribution $P$ that allows for
a density $p$ with respect to the Lebesgue measure and
is generated by model~(\ref{SEMadd}) with DAG $D^0$ and nonlinear,
three times differentiable functions $f_{j,k}^0$.
Assume further the condition from Lemma~\ref{lemclosed}.
Then $\xi_p > 0$.
%Assume further\vspace*{-2pt} that
%$\int\frac{p(x_j,x_k)^2}{p(x_j)p(x_k)} \,dx_j \,dx_k < \infty
%$ for any $j,k \in\{1, \ldots, p\}$. Then $\xi_p > 0$.
\end{lemm}
%
%Vllt. koennen einige der Lemma in den Appendix.}

\begin{pf}
Because of the closedness of $\mathcal{F}^{\oplus
j}$ (Lemma~\ref{lemclosed}), the minimum in~(\ref{eqproje}) is obtained
for some functions $f_{j,k}$. Without loss of generality, we can assume
that all constant additive
components are zero.
%(if $\EE[X_j]$ would not be zero, we would include an
%intercept in the corresponding additive structural equation).
%Widerspruch zu Lemma1. Bekommen wir den noch, falls die $f_{j,k}$s
%konstant sind? Oder ist das irgendwie ersichtlich aus~
But then $\xi_p=0$ would contradict Lemma~\ref{lemid}.
\end{pf}

The number $\xi_p$ describes the degree of separation
between the true model and misspecification when using a wrong
permutation. As discussed in Remark~\ref{rem1}, $\xi_p = 0$ for the case
of linear Gaussian SEMs.
Formula (\ref{identifiability}) can be expressed as
%e9 #&#
%
\begin{equation}
\label{identif2} %& &\min_{\pi\notin\Pi^0}p^{-1} \sum_{j=1}^p
%(\log(\sigma^{\pi,0}_{\pi^{-1}(j)}) - \log(\sigma_{j}^0))\nonumber\\
\xi_p = \min
_{\pi\notin\Pi^0}p^{-1} \sum_{j=1}^p
\bigl(\log\bigl(\sigma^{\pi,0}_{j}\bigr) - \log\bigl(
\sigma_{j}^0\bigr)\bigr) \ge0.
\end{equation}
%
%where typically $\xi_p > 0$. \Jonas{What does ``typically'' mean?
%Ausserdem verstehe ich diese Gleichung nicht -fuerchte ich.}
%The number $\xi_p$ describes the degree of separation
%between the true model and misspecification when using a wrong
%permutation. As discussed in Remark~\ref{rem1}, $\xi_p = 0$ for the
%case
%of linear Gaussian SEMs.

%re2 #&#
%
\begin{rem}\label{remark2}
Especially for situations where $p$ is very large so that the factor
$p^{-1}$ is small, requiring a lower bound
$\xi_p > 0$ can be overly restrictive. Instead of requiring a gap with the
factor $p^{-1}$ between the likelihood scores of the true distribution
and all distributions corresponding to permutations, one can weaken
this as
follows. Consider $H(D,D^0) = \{j; \pa_{D^0}(j) \nsubseteq\pa_D(j)\}
$. We
require that
%e10 #&#
%
\begin{equation}
\label{weakenxi} \xi_p':= \min_{D \neq D^0}
\bigl\llvert H\bigl(D,D^0\bigr)\bigr\rrvert^{-1} \sum
_{j \in H(D,D^0)} \bigl(\log\bigl(\sigma_j^{D,0}
\bigr) - \log\bigl(\sigma_j^0\bigr)\bigr) \ge0,
\end{equation}
where\vspace*{1pt} $(\sigma_j^{D,0})^2$ is the error variance in the best additive
approximation of $X_j$ based on $\{X_k; k \in\pa_D(j)\}$. Such a
weaker gap
condition is proposed in \cite{lohbu13}, Section~5.2. All our theoretical
results still hold when replacing statements involving $\xi_p$ in
(\ref{identif2}) by the corresponding statements with $\xi'_p$ in
(\ref{weakenxi}).
\end{rem}

%s2.4 #&#
\subsection{Maximum likelihood estimation for order: Low-dimensional setting}\label{subsecmle}

%For estimation we want to achieve/show that
We assume having $n$ i.i.d. realizations $X^{(1)},\ldots,X^{(n)}$ from model
(\ref{modsem}). For\vspace*{1pt} a $n \times1$ vector
$x = (x^{(1)},\ldots,x^{(n)})^T$, we denote by
$\llVert x\rrVert _{(n)}^2 = n^{-1} \sum_{i=1}^n (x^{(i)})^2$.
Depending on the
context, we sometimes denote by $\hat{f}$ a function and sometimes an $n
\times1$ vector evaluated at (the components of) the data points $X^{(1)},\ldots,X^{(n)}$; and similarly
for~$X^{\pi}_j$.
We consider the unpenalized maximum likelihood estimator:
\begin{eqnarray*}
\hat{f}_{j}^{\pi} &=& \argmin_{g_j \in{\mathcal F}_n^{\oplus j-1}}
\Biggl\llVert
X^\pi_j - \sum_{k=1}^{j-1}
g_{j,k}\bigl(X^{\pi}_k\bigr)\Biggr\rrVert
_{(n)}^2,
\qquad
\bigl(\hat{\sigma}^{\pi}_j\bigr)^2 = \Biggl
\llVert X^\pi_j - \sum_{k=1}^{j-1}
\hat{f}_{j,k}^{\pi}\bigl(X^{\pi}_k\bigr)
\Biggr\rrVert_{(n)}^2.
\end{eqnarray*}
Denote by $\hat{\pi}$ a permutation which minimizes the
unpenalized negative log-likelihood:
%e11 #&#
%
\begin{equation}
\label{mlepi} \hat{\pi} \in\argmin_{\pi} \sum
_{j=1}^p \log\bigl(\hat{\sigma}^{\pi}_j
\bigr).
\end{equation}

Estimation of $\hat{f}_j^{\pi}$ is based on ${\mathcal F}_n$ with
pre-specified
basis functions $b_r(\cdot)$ with $r = 1,\ldots,a_n$. In practice, the
basis functions could depend on the predictor variable or on the order of
variables, for example, when choosing the knots in
regression splines. The classical choice for the number of basis functions
is $a_n \asymp n^{1/5}$ for twice differentiable functions: here, and
as explained in Section~\ref{secconsistency}, however, a~smaller
number such as $a_n = O(1)$ to detect some nonlinearity might be
sufficient for estimation of the true underlying order.
% you try to use a few basis functions only? See also my comment in
%Section
% 5.2}

%leaves room for improvement. What about: \\
%(1) $\hat\pi$ for the estimate for the permutation (as before).\\
%(2) $D^{\pi}_{full}$ for the full DAG corresponding to the order $\pi$
%(as before).\\
%(3a) $S(D^{\pi}_{full})$ for the corresponding DAG after (correct)
%variable selection (right now I think it only exists as $S^{0}$ for $
%(3b) $\hat S(D^{\pi}_{full})$ for the corresponding DAG after variable
%selection (instead of $\hat S^{\pi}$). \\
%(4) $D^{\pi}_{restr}$ for the DAG from section~\ref{secrestrmle}. \\
%And (5) $S(D^{\pi}_{restr})$ for the corresponding version after
%variable selection.\\
%This stresses the two step procedure (first full DAG, then variable
%selection).}
%see below???}

%s2.5 #&#
\subsection{Sparse regression for feature selection}\label{subsecvarsel}

Section~\ref{secconsistency} presents assumptions and results ensuring
that with high probability $\hat{\pi} = \pi^0$ for some $\pi^0 \in
\Pi^0$.
With such an estimated order $\hat{\pi}$, we obtain a complete super-DAG
(super-graph) $D^{\hat\pi}$ of the underlying DAG $D^0$ in~(\ref{modsem}),
where the parents of a node $\hat{\pi}(j)$ are defined as
$\pa_{D^{\hat{\pi}}}(\hat{\pi}(j)) = \{\hat{\pi}(k); k<j\}$
for all $j$.
We can pursue consistent
estimation of intervention distributions based on $D^{\hat\pi}$
without any additional need to
find the true underlying DAG $D^0$; see Section~\ref{subseccausalconsist}.

However, we can improve statistical efficiency for estimating the intervention
distribution when it is ideally based on
the true DAG $D^0$ or realistically a not
too large super-DAG $\hat{D}^{\hat{\pi}} \supseteq D^0$.
The task of estimating such a super-DAG $\hat{D}^{\hat{\pi}}
\supseteq
D^0$ is conceptually straightforward: starting from the complete super-DAG
$D^{\hat\pi}$ of $D^0$ as discussed above,
we can use model selection or a penalized multivariate
(auto-) regression technique in the model representation (\ref{triangular}).
%}
%without too much additional notation}.
For additive model fitting, we can either use hypothesis testing for
additive models \cite{marrawood11} or
%look
%at p-values. Since we never use group lasso, we might want to adjust
%this...} we can use
the Group Lasso~\cite{ravik09}, or its improved version with a
sparsity-smoothness penalty proposed in
\cite{meieretal09}. All the techniques mentioned above perform variable
selection, where we denote by
\[
\hat{D}^{\hat{\pi}} = \bigl\{\bigl(\hat{\pi}(k),\hat{\pi}(j)\bigr);
\hat{f}_{j,k}^{\hat{\pi}} \not\equiv c\bigr\},
\]
(the constant $c = 0$ when assuming that $\hat{f}_{j,k}^{\hat{\pi}}$ have
mean zero when evaluated over all data-points)\vspace*{2pt}
the selected variables indexed in the original order [we
obtain estimates $\hat{f}_{j,k}^{\hat{\pi}}$ in the representation\vspace*{-1pt}
(\ref{triangular}) with correspondence to the indices
$\hat{\pi}(k),\hat{\pi}(j)$ in the original order]; we identify
these selected
variables in $\hat{D}^{\hat{\pi}}$ as the edge set of~a~DAG.
For example, with the Group Lasso, assuming some condition avoiding near
collinearity of functions, that is,
a compatibility condition for the Group
Lasso \cite{pbvdg11}, Chapter~5.6, Theorem~8.2, and that the $\ell_2$-norms
of the nonzero
functions are sufficiently large,\vspace*{1.5pt} we obtain the screening property (since
we implicitly assume that $\hat{\pi} \in \Pi^0$ with high probability):
with high probability and asymptotically
tending to one,
%e12 #&#
%
\begin{equation}
\label{screening} \hat{D}^{\hat{\pi}} \supseteq D^0 = \bigl\{(k,j);
f^0_{j,k} \not\equiv0\bigr\}
\end{equation}
saying that all relevant variables (i.e., edges) are selected.
%switching $j$ and $k$: $\hat{S}^{\pi^0} = \{(\pi^0(k),\pi^0(j))); %
%contains
%now all edges after feature selection.}\Peter{Done.}
Similarly with hypotheses testing, assuming that the nonzero
$f^0_{j,k}$ have\vspace*{1pt}
sufficiently large $\ell_2$-norms, we also obtain that (\ref
{screening}) holds
with high probability.

The same argumentation applies if we use $D^{\hat\pi}_{\mathrm{restr}}$
from Section~\ref{secrestrmle} instead of $D^{\hat\pi}$ as an initial
estimate. This then results in $\hat D^{\hat\pi}_{\mathrm{restr}}$,
replacing $\hat D^{\hat\pi}$ above.

%I think we could do it without, this could save a subsubscript...}
% \Peter{Yes: because we ``assume'' the idealistic setting where we have
% found a true permutataion $\pi^0$}
% \Jonas{I guess, my idea was to allow for a notation for our final DAG
% estimate. It should be $\hat D^{\hat\pi}$, shouldn't it? (Or, $\hat
% D^{\hat\pi}_{restr}$ if we include prel. neighb. sel.) As far as I
%can
% tell, this never appears anywhere in the paper. But -for me- the
% statement that this estimate $\hat D^{\hat\pi}$ (or even $\hat D^{
% \pi}_{restr}$) leads to consistent estimation of the causal effects is
% THE theoretical result of the paper. Or am I missing sth?}

%s2.6 #&#
\subsection{Consistent estimation of causal effects}\label{subseccausalconsist}
% \Jonas{This section I would really like to be 2.5 instead of 2.4.1.
%For me, it is not a subproblem or substh. of penalized regression. It
%rather concerns both dags $D^{\pi^0}_{full}$ and $\hat D^{\pi^0}$,
%whereas only the latter belongs into 2.4.}
%The screening property \eqref{screening} for regression immediately
%translates to estimation of the DAG. From $\hat{S}^{\pi^0}$ we obtain
%the
%DAG $\hat{D}^{\pi^0}$, by defining presence of an edge $k \to j$ in
%the DAG
%if and only if $(k,j) \in\hat{S}^{\pi^0}$.
%suggested notation above)}
%%\Jonas{Aha, without the change above, there might be a small mistake
%%here. Should be checked at least}:
%%Peter: Done
%The DAGs $\hat{D}^{\pi^0}$ might be
%different for various $\pi^0 \in\Pi^0$. If the true DAG $D^0$ is
%identifiable from the distribution,
%e.g. in an additive structural equation models as in \eqref{modsem},
%all
%of these estimated DAGs satisfy:
The property in (\ref{screening}) has an important implication for causal
inference:\footnote{We assume that interventions at variables do not change
the other structural equations, and that there are no unobserved hidden
(e.g., confounder) variables.
% \Peter{Siehe Kommentar 17 vom referee: ist das OK so?}
} all estimated causal effects and estimated
intervention
distributions based on the estimated DAG $\hat{D}^{\hat{\pi}}$ are
consistent. In fact, using the do-calculus \cite{pearl00}, cf. (3.10),
we have for the
single intervention (at variable $X_k$) distribution for $X_j$, for all $j
\neq k$:
\[
p_{D^0}\bigl(x_j\mid \do(X_k
= x)\bigr) = p_{\hat{D}^{\hat{\pi
}}}\bigl(x_j\mid
\do(X_k = x)\bigr)\qquad\mbox{for all } x,
\]
where $p_{D}(\cdot\mid\do(\cdot))$ denotes the intervention density
based on a DAG $D$.
%from
% the permutation, right? One could make the argument that after the
% pruning, the interventional distributions are easier to estimate?}

We note that the screening property (\ref{screening}) also holds when
replacing $\hat{D}^{\hat{\pi}}$ with
the full DAG induced by $\hat{\pi}$, denoted by $D^{\hat{\pi}}$.
Thus, the feature selection step in Section~\ref{subsecvarsel} is not
needed to achieve consistent estimation of
causal effects. However, a
smaller DAG $D^0 \subseteq\hat{D}^{\hat{\pi}} \subseteq
D^{\hat{\pi}}$ typically leads to better (more statistically efficient)
estimates of
the interventional distributions than the full DAG
$D^{\hat{\pi}}$.

\section{Restricted maximum likelihood estimation: Computational and statistical benefits}

We present here maximum likelihood estimation where we restrict the
permutations, instead of searching over all permutations in
(\ref{mlepi}). Such a restriction makes the computation more tractable,
and it is also statistically crucial when dealing with high-dimensional
settings where $p > n$.

%s3.1 #&#
\subsection{Preliminary neighborhood selection} \label{secrestrMLE}

We first perform neighborhood selection with additive models, following the
general idea in \cite{mebu06} for the linear Gaussian case.
%the
%OR rule. Since being a possible parent is directed, we do not combine
%the
%results for $k$ on $j$ and $j$ on $k$. (2) as described in the method
%section, we are using gamboost rather than group lasso. but maybe we do
%not need to stress that here.}
We pursue variable
selection in an additive model of $X_j$ versus all other variables
$X_{\{-j\}} = \{X_k; k \neq j\}$: a natural
method for such a feature
selection is the Group Lasso for additive models \cite{ravik09},
ideally with
a sparsity-smoothness penalty~\cite{meieretal09}; see also
\cite{vooretal13}. This provides us with a set of variables
\[
\hat{A}_j \subseteq\{1,\ldots,p\} \setminus j
\]
which denotes the selected variables in the estimated conditional expectation
\[
\hat{\EE}_{\mathrm{add}}[X_j\mid X_{\{-j\}}] = \sum
_{k \in\hat{A}_j} \hat{h}_{jk}(X_k)
\]
with functions
$\hat{h}_{jk}$ satisfying $n^{-1} \sum_{i=1}^n \hat
{h}_{jk}(X_k^{(i)}) =
0$ (i.e., a possible intercept is subtracted already): that is,
\[
\hat{A}_j = \{k; k \neq j, \hat{h}_{j,k} \not\equiv0\}.
\]
%
%doch gar nicht. Stattdessen kommt das Folgende: OK?}
We\vspace*{1pt} emphasize that the functions
$\hat{h}_{j,k}(\cdot)$ are different from $\hat{f}_{j,k}^{\pi
}(\cdot)$ in
Section~\ref{subsecmle} because for the former, the additive
regression is
against all other variables.
%$\hat{D}^{\pi^0}$
%in Section~\ref{subsecvarsel} because for the former, the additive
%regression is
%against all other variables; similarly, the functions
%$\hat{h}_{j,k}(\cdot)$ are different from $\hat{f}_{j,k}^{\pi}(\cdot)$
%in
%Section~\ref{subsecmle}.

%We define a directed graph $\hat{G}$ with an edge $k \rightarrow j$ if
%and only
%if $k \in\hat{A}_j$.
%%\Peter{Wieso nicht undirected? Siehe Kommentar 18. vom referee.}
%%\Jan{Ich denke, directed ist hier die richtige Aussage. $k$ kann ein
%possible %parent
%% von j sein, ohne dass j ein possible parent von k ist}
%%The union of
%%the selected variables is denoted by
%%\begin{eqnarray*}
%%\hat{A} = \cup_{j=1}^p \hat{A}_j.
%%\end{eqnarray*}
%We emphasize that the edge set in $\hat{G}$ is different from
%$\hat{D}^{\pi^0}$
%in Section~\ref{subsecvarsel} because for the former, the additive
%regression is
%against all other variables; similarly, the functions
%$\hat{h}_{j,k}(\cdot)$ are different from $\hat{f}_{j,k}^{\pi}(\cdot)$
%in
%Section~\ref{subsecmle}.

We give conditions in Section~\ref{subsecrmletheory} (see Lemma~\ref{lemm0})
ensuring that the neighborhood selection
set contains the parental variables from the structural equation model in
(\ref{SEMadd}) or (\ref{modsem}), that is, $\hat{A}_j \supseteq
\pa(j)$.

%s3.2 #&#
\subsection{Restricted maximum likelihood estimator} \label{secrestrmle}

We restrict the space of permutations in the definition of (\ref{mlepi})
such that they are ``compatible'' with the neighborhood selection sets
$\hat{A}_j$. Note that for the estimator $\hat{\sigma}^{\pi}_j$ in
(\ref{mlepi}), we regress $X_{\pi(j)}$ against $\{X_k; k \in
\{\pi(j-1),\ldots, \pi(1) \}\}$. We restrict here the set of
regressors to
the indices
$R_{\pi,j} =\{\pi(j-1),\ldots, \pi(1)\} \cap\hat{A}_{\pi(j)}$.
We then
calculate the $\pi(j)$th term of the
log-likelihood using the set of regressors $X_{R_{\pi,j}} = \{X_k; k
\in
R_{\pi,j}\}$.
%A permutation $\pi$ induces a full DAG $D_{\mathrm{full}}^{\pi}$ where
%$\pa_{D_{\mathrm{full}}^{\pi}}(j) = \{\pi(k); k<j\}$ for all $j$.
%Consider
%R_j = \{\pi; \pa_{D_{\mathrm{full}}^{\pi}\right\vert_{\hat{A}}}(j)
%where $D\left\vert_{\hat{A}}$ denotes the restriction of DAG $D$ to
%the set of
%variables $\hat{A}$.
%nicht gesagt%, dass jedes j in einem $\hat A_k$ vorkommt, oder? Somit
%waere j auch nicht not%wendiger Weise in $\hat A$. Und dann verstehe
%ich die Aussage ``parents von j i%m DAG eingeschraengt auf $\hat A$''
%nicht. Kann es sein, dass man die Einschrae%nkung einfach weglassen
%kann?}
%Thus, $R_j$ includes all permutations whose induced
%parents, for the node $j$, are included in $\hat{A}_j$. The overall
%restriction is
%R = \cap_{j=1}^p R_j,
%and the restricted maximum likelihood estimator is
More precisely, we estimate
\begin{eqnarray*}
\hat{f}_{j}^{\pi,R} &=& \argmin_{g_{j,k} \in \mathcal{F}_n}\Biggl\|X^\pi_j
- \sum_{k; \pi(k) \in R_{\pi,j}}
g_{j,k}\bigl(X^{\pi}_k\bigr)\Biggr\rrVert
_{(n)}^2,
\\
\bigl(\hat{\sigma}^{\pi,R}_j\bigr)^2 &=& \biggl
\llVert X^\pi_j - \sum_{k; \pi(k)
\in R_{\pi,j}}
\hat{f}_{j,k}^{\pi,R}\bigl(X^{\pi}_k\bigr)
\biggr\rrVert_{(n)}^2,
\end{eqnarray*}
and the restricted maximum likelihood estimator is
%e13 #&#
%
\begin{equation}
\label{rmlepi} \hat{\pi} \in\argmin_{\pi} \sum
_{j=1}^p \log\bigl(\hat{\sigma}^{\pi,R}_j
\bigr).
\end{equation}
%
%where
%& &\hat{f}_{j}^{\pi,\hat{A}_j} = \argmin_{g_j \in{\mathcal F}_n^{
%- \sum_{k \in\hat{A}_{\pi(j)}} g_{j,k}(X^{\pi}_k)\right\Vert_{(n)}^2,\
%& &(\hat{\sigma}^{\pi,\hat{A}_{\pi(j)}}_j)^2 = \left\Vert X^\pi_j -
%are restricted to the feature selection sets $\hat{A}_{\pi(j)}$.
If $\max_j\llvert \hat{A}_j\rrvert < n$, the estimators
$\hat{\sigma}_j^{\pi,R}$ are well defined.

The computation of the restricted maximum likelihood estimator in
(\ref{rmlepi}) is substantially easier than for the unrestricted MLE
(\ref{mlepi}) if $\max_j\llvert \hat{A}_j\rrvert $ is small (which is
ensured if
the true
neighborhoods are sparse).
%That is, define the equivalence relation
%R_{\pi',\pi'^{-1}(j)} \forall j.
%where $c(j,\pi,\pi') = \pi^{'-1} \pi(j)$, i.e., $c(j,\pi,\pi')=:j^*$
%is a
%renumbering of the index $j$ such that $\pi' (j^*) = \pi(j)$.
%Then,
The set of all permutations can be partitioned in equivalence
classes $\bigcup_r {\mathcal R}_r$ and the minimization in (\ref
{rmlepi}) can be
restricted to single
representatives of each equivalence class ${\mathcal R}_r$. The equivalence
relation
%above might be more intuitive when using the notion of a
can be formulated with a
restricted DAG $D_{\mathrm{restr}}^{\pi}$ whose parental set for node
$\pi(j)$
equals $\pa_{D_{\mathrm{restr}}^{\pi}}(\pi(j)) = R_{\pi,j}$. We
then have that
\[
\pi\sim\pi'\quad\mbox{if and only if}\quad D_{\mathrm{restr}}^{\pi} =
D_{\mathrm{restr}}^{\pi'}.
\]

%koennen. Ist ja nur ne zusaetzliche Permutation der Indizes. Also\dvtx
%Zudem\dvtx  Wenn wir den zugehoerigen Graphen $D_{restr}^{\pi}$ definieren
%(s.o.), dann koennen wir einfach sagen
%Ueberlegung\dvtx  Kann man $R_{\pi,j}$ so umdefinieren, dass es die Parents
%von $j$ fuer Order $\pi$ sind (und nicht\dvtx  $R_{\pi,\pi^{-1}(j)}$ sind
%die Eltern von $j$)? Ich glaube, oben bei der Definition einfach $R_{
%}
%meiner Meinung nach auch
%schreiben. Ich finde das etwas einleuchtender, aber nicht unbedingt
%alle. In an%y case, wir sollten uns vllt. fuer $\sim$ oder $\equiv$
%entscheiden.
%}

%$\{j-1,\ldots, 1\}
%The latter is ensured if
%the true underlying neighborhoods are sparse.
Computational details are described in Section~\ref{seccomputation}.

%s4 #&#
\section{Consistency in correct and misspecified models}\label{secconsistency}

We prove consistency for the ordering among variables in
additive structural equation models, and under an additional identifiability
assumption even for the case where the model is misspecified with
respect to the error distribution or when using highly biased function
estimation.
%one
%more assumption than the correctly specified one, or not?}

%s4.1 #&#
\subsection{Unrestricted MLE for low-dimensional settings}\label{subsecmletheory}

We first consider the low-dimensional setting where $p< \infty$ is fixed
and $n \to\infty$, and we establish consistency of the unrestricted
MLE in
(\ref{mlepi}).
We assume the following:
\begin{longlist}[(A4)]
\item[(A1)] Consider a partition of the real line
\[
\R= \bigcup_{m=1}^{\infty} I_m
\]
using disjoint intervals $I_m$. The individual functions in ${\mathcal
F}$ are
$\alpha$-times differentiable, with $\alpha\ge1$, whose derivatives
up to
order $\alpha$ are bounded in absolute value by $M_m$ in $I_m$.
%This ensures
% some constants $0 < C_1,C_2 < \infty$, independent of $g$,
%h_{j,g}^{\pi}(X_i) - \EE[h_{j,g}^{\pi}(X_i)]\left\vert> x] \le C_1
%x), %j=2,\ldots,p.
%
\item[(A2)] Tail and moment conditions:
\begin{enumerate}[(ii)]
\item[(i)] For $V = 1/\alpha$ and $M_m$ as in (A1):
\[
\sum_{m=1}^{\infty} \bigl(M_m^2
\PP[X_j \in I_m]\bigr)^{V/(V+2)} < \infty,\qquad j=1,\ldots,p.
\]
\item[(ii)]
\begin{eqnarray*}
\EE\llvert X_j\rrvert^4 &<& \infty,\qquad j=1,\ldots,p,
\\
\sup_{f \in{\mathcal F}} \EE\bigl\llvert f(X_j)\bigr\rrvert
^{4} &<& \infty,\qquad j=1,\ldots,p.
\end{eqnarray*}
\end{enumerate}
\item[(A3)] The error variances satisfy $(\sigma_{j}^{\pi,0})^2 > 0$ for
all $j=1,\ldots,p$ and all $\pi$.
\item[(A4)] The true functions $f_{j,k}^0$ can be approximated on any
compact set ${\mathcal C} \subset\R$: for all $k \in
\pa_{D^0}(j), j=1,\ldots,p$,
\[
\EE\bigl[\bigl(f_{j,k}^0(X_k) -
f_{n;j,k}^0(X_k)\bigr)^2
I(X_k \in{\mathcal C})\bigr] = o(1),
\]
where
\[
f_{n;j}^0 = \argmin_{g_j \in \mathcal{F}_n^{\oplus j-1}} \EE\Biggl[\Biggl(X_j -
\sum_{k \in \pa_{D^0}(j)} g_{j,k}(X_k)\Biggr)^2\Biggr].
\]
%
%Furthermore, for some $\kappa> 0$, $\EE\left\vert f_{j,k}^0(X_k)\left
%< \infty%$ and
%$\EE\left\vert f_{n;j,k}^0(X_k)\left\vert^{2 + 2\kappa} < \infty$ for
%all $k=1,\ldots
%,j-1$
%and $j = 2,\ldots,p$.
\end{longlist}
All assumptions are not very restrictive. The second part of assumption
\textup{(A2)(ii)} holds if we assume, for example, a bounded function class
${\mathcal
F}$, or if $\llvert f(x)\rrvert \asymp\llvert x\rrvert $ as
$\llvert x\rrvert \to\infty$ for all $f \in
{\mathcal
F}$.

%th1 #&#
%
\begin{theo}\label{th1}
Consider an additive structural equation model as in
(\ref{modsem}). Assume \textup{(A1)--(A4)} and
$\xi_p > 0$ in (\ref{identifiability}) (see also Lemma~\ref{lemxip}
and Remark~\ref{remark2}). Then we have
\[
\PP\bigl[\hat{\pi} \in\Pi^0\bigr] \to1\qquad (n \to\infty).
\]
\end{theo}

A proof is given in the supplemental article \cite{CAMsuppl14}. %
%As mentioned before, the assumption $\xi_p > 0$ holds if all functions
%$f_{j,k}^0$ are nonlinear \cite{petersetal13}.
%Theorem~\ref{th1} says that for low- and medium-dimensional problems
%where
%the dimension is allowed to grow like $p = o((n/\log^2(n))^{1/(2 + 1/
%assuming that $\xi_p$ is of order one,
Theorem~\ref{th1} says that one can find a correct order among the
variables without pursuing feature or edge selection for the structure in
the SEM.
%Note that for very smooth functions, the growth in
%dimensionality is of the order $p = o(n^{1/2 - \kappa})$ with
%$\kappa> 0$ arbitrarily small.

%re3 #&#
%
\begin{rem}\label{remark3}
Studying near nonidentifiable models, for example, near
linearity in a Gaussian structural equation model, can be modelled by allowing
$\xi_p = \xi_{n,p}$ to converge to zero as $n \to \infty$. If
one requires $\xi_{n,p} \gg n^{-1/2}$, the statement of Theorem~\ref{th1}
still holds. We note that Theorem~\ref{th3} for the high-dimensional case
implicitly allows $\xi_p = \xi_{p_n}$ to change with sample size $n$.
However, it is a nontrivial issue to translate such a
condition in terms of closeness of one or several nonlinear
functions $f_{j,k}^0$ to their closest linear approximations.
Similarly, if
some error variances $\sigma_j^{\pi,0}$ would be close to zero
(e.g., converge to zero as $n \to\infty$ asymptotically), this could cause
identifiability problems such that
$\xi_p$ might be close to (e.g., converge fast to) zero.
%Such issues are vaguely related to near
%nonidentifiability questions for the Markov equivalence classes in
%linear
%Gaussian structural equation model \cite{uhleretal13}.
\end{rem}

Related to Remark~\ref{remark3} is the question about uniform convergence
in the statement of Theorem~\ref{th1}, over a whole class of structural
equation models. This can be ensured by strengthening the assumptions to
hold uniformly:
\begin{longlist}[(U4)]
\item[(U1)] The quantities in \textup{(A2)(i)} and \textup{(ii)} are upper-bounded by
positive constants
$C_1 < \infty, C_2 < \infty$ and $C_3 < \infty$.
\item[(U2)] The error variances in \textup{(A3)} are lower bounded by a finite
constant $L > 0$.
\item[(U3)] The approximation in \textup{(A4)} holds uniformly over a class of functions
${\mathcal F}$: for any compact set ${\mathcal C}$ and any $j,k$:
\[
\sup_{f^0 \in{\mathcal F}} \EE\bigl[\bigl(f^0_{j,k}(X_k)
- f^0_{n;j,k}(X_k)\bigr)^2
I(X_k \in{\mathcal C})\bigr] = o(1).
\]
\item[(U4)] The constant $\xi_p \ge B > 0$ for some finite constant
$B >0$.
\end{longlist}
Denote the class of distributions in an additive SEM which satisfy
\textup{(U1)--(U4)} by ${\mathcal P}(C_1,C_2,L,{\mathcal F},B)$. We then obtain a
uniform convergence result
%e14 #&#
%
\begin{equation}
\label{uniform} \inf_{P \in{\mathcal P}(C_1,C_2,C_3,{\mathcal F},L)} \PP
_P\bigl[\hat{\pi}
\in\Pi^0\bigr] \to1\qquad (n \to\infty).
\end{equation}
This can be shown exactly along the lines of the proof of Theorem
\ref{th1} in the supplemental article \cite{CAMsuppl14}.

%s4.1.1 #&#
\subsubsection{\hspace*{-1pt}Misspecified error distribution and biased function estimation} \label{secmissp}
\hspace*{-2pt}Theorem~\ref{th1} generalizes to the situation where the model in
(\ref{modsem}) is misspecified and the truth has independent, non-Gaussian
errors $\varepsilon_1,\ldots,\varepsilon_p$ with $\EE[\varepsilon
_j] = 0$. As in Theorem
\ref{th1}, we make the assumption $\xi_p > 0$ in (\ref{identif2}): its
justification, however, is somewhat
less backed up because the identifiability results from
\cite{petersetal13} and Lemma~\ref{lemxip} do not carry over
immediately. The latter results say that the set
of correct orderings $\Pi^0$ can be identified from the distribution of
$X_1,\ldots,X_p$, but we require in (\ref{identif2}) that identifiability
is given in terms of all the error variances, that is, involving only second
moments. It is an open problem whether (or for which subclass of models)
identifiability from the distribution carries over to automatically ensure
that $\xi_p > 0$ in~(\ref{identif2}).
%and we require here that identifiability is given by a
%function of all the
%error variances $\sigma_j^2 (j=1,\ldots,p)$, as for the case of
%Gaussian errors; it is an open problem whether (or for which subclass
%of models) identifiability from the
%distribution carries over to automatically ensure that $\xi_p > 0$ in
%formulieren?}
%Then, the best projected
%model with Gaussian errors
%(w.r.t. Kullback-Leibler divergence) would
%still allow for identifiability, as a consequence of the result in
%identifiability in the best projected model with Gaussian errors.
%stil%l do not understand this projection. (1) Is it obvious that (when?)
%the projecte%d model is again nonlinear? (2) What happens if the
%distribution is closer to t%he model class with a wrong
%permutation????}

Furthermore, assume that the number of basis functions $a_n$ for functions
in ${\mathcal F}_n$ is small such
that assumption \textup{(A4)} does not hold, for example, $a_n = O(1)$. We
denote by
\[
\bigl(\sigma_j^{\pi,0,a_n}\bigr)^2 = \min
_{g_j \in{\mathcal F}_n^{\oplus
j-1}}\EE_{\theta^0}\Biggl[\Biggl(X^{\pi}_j
- \sum_{k=1}^{j-1} g_{j,k}
\bigl(X^{\pi}_k\bigr)\Biggr)^2\Biggr],
\]
which is larger than $(\sigma_j^{\pi,0})^2$ in (\ref{add1}). Instead of
(\ref{identif2}), we then consider
%e15 #&#
%
\begin{equation}
\label{identif3} \xi_{p}^{a_n}:= \min_{\pi\notin\Pi^0, \pi^0 \in\Pi^0}p^{-1}
\sum_{j=1}^p \bigl(\log\bigl(
\sigma^{\pi,0,a_n}_{j}\bigr) - \log\bigl(\sigma_{j}^{\pi^0,0,a_n}
\bigr)\bigr).
\end{equation}
Requiring
\[
\liminf_{n \to\infty}\, \xi_{p}^{a_n} > 0
\]
is still reasonable: if (\ref{identif2}) with $\xi_p > 0$ holds
because of
nonlinearity of the additive functions \cite{petersetal13}, and see the
interpretation above for non-Gaussian errors, we believe that it typically
% maybe my question is about this ``typically''. For me, it would also
%be
% fine to leave it this way.}
also holds for the best projected additive functions in $\mathcal{F}_n^{\oplus}$ as long as some nonlinearity is present when
using $a_n$ basis functions; here, the best projected additive function\vspace*{-1pt} for
the $j$th
variable $X^{\pi}_j$ is defined as $f_{n;j}^{\pi} = \operatorname{argmin}_{g_j \in
{\mathcal
F}_n^{\oplus j-1}} \EE[(X^{\pi}_j - \sum_{k=1}^{j-1}
g_{j,k}(X^{\pi}_k))^2]$.
We also note that for $a_n \to\infty$, even when diverging very
slowly, and assuming \textup{(A4)} we have that $\xi_p^{a_n} \to\xi_p$ and thus
$\liminf_{n \to\infty} \xi_{p}^{a_n} > 0$.
In general, the choice of the number of basis functions $a_n$ is a trade-off
between identifiability (due to nonlinearity) and
estimation accuracy: for $a_n$ small we might have a smaller value in
(\ref{identif3}), that is, it might be that $\xi_p^{a_n} \le\xi
_p^{a'_n}$ for
$a_n \le a'_n$, which makes identifiability harder but exhibits less
variability in
estimation, and vice versa. In particular, the trade-off between
identifiability and variance might be rather different than the classical
bias-variance trade-off with respect to prediction in classical function
estimation. A~low complexity (with $a_n$ small) might be better than a
prediction optimal number of basis functions.
% we will exploit this fact in
%Section~\ref{secnumerical}.\Peter{Ist das gemacht?}
%SID resultaten einfach den RSS vom gam fitting im WAHREN graphen
%plotten.}

Theorem~\ref{th2} below establishes the consistency for order estimation
in an additive structural equation model with potentially non-Gaussian
errors, even when the expansion for function estimation is truncated
at few basis functions.

%th2 #&#
%
\begin{theo}\label{th2}
Consider an additive structural equation model as in (\ref{modsem})
but with
independent potentially non-Gaussian errors $\varepsilon_1,\ldots,\varepsilon_p$ having\break
$\EE[\varepsilon_j] = 0$ $(j=1,\ldots,p)$. Assume either of the following:
\begin{longlist}[1.]
\item[1.]\textup{(A1)--(A4)} hold, and $\xi_p > 0$ in formula (\ref{identif2})
(see also
Remark~\ref{remark2}).
\item[2.]\textup{(A1)--(A3)} hold, and $\liminf_{n \to\infty} \xi_{p}^{a_n} >0$ in
formula (\ref{identif3}).
\end{longlist}
Then
\[
\PP\bigl[\hat{\pi} \in\Pi^0\bigr] \to1\qquad (n \to\infty).
\]
\end{theo}

A proof is given in the supplemental article \cite{CAMsuppl14}. Again,
as appearing in the discussion of
Theorem~\ref{th1}, one can obtain uniform convergence by strengthening the
assumptions to hold uniformly over a class of distributions.

%s4.2 #&#
\subsection{Restricted MLE for sparse high-dimensional setting}\label{subsecrmletheory}

We consider here the restricted MLE in (\ref{rmlepi}) and show that
it can
cope with high-dimensional settings where $p \gg n$.

The model in (\ref{SEMadd}) is now assumed to change with sample size
$n$: the
dimension is $p = p_n$ and the parameter $\theta= \theta_n$ depends on
$n$. We consider the limit as $n \to\infty$ allowing diverging dimension
$p_n \to\infty$ where $p_n \gg n$. For notational simplicity, we often
drop the sub-index $n$.

We make a few additional assumptions.
%In contrast to the situation from Section~\ref{subsecmletheory}, we
%make
%assumptions which exclude concurvity (or collinearity) among the
%additive
%functions.
When fitting an additive model of $X_j$ versus all other variables
$X_{\{-j\}}$, the target of such an estimation is the best approximating
additive function:
\begin{eqnarray*}
\EE_{\mathrm{add}}[X_j\mid X_{\{-j\}}] &=& \sum
_{k \in\{-j\}} h_{jk}^*(X_k),
\\
\bigl\{h_{jk}^*; k \in \{-j\}\bigr\} &=& \argmin_{h_j \in \mathcal{F}^{\oplus
p-1}}
\EE\biggl[\biggl(X_j - \sum_{k \in \{-j\}} h_{jk}(X_k)\biggr)^2\biggr].
\end{eqnarray*}
In general, some variables are irrelevant, and we denote the set of relevant
variables by $A_j$: $A_j \subseteq\{1,\ldots,p\} \setminus j$ is the (or
a) smallest set\footnote{Uniqueness of such a set is not a requirement but
implicitly ensured by the compatibility condition and sparsity which we
invoke to guarantee \textup{(B2)(ii)}.}
such that
\[
\EE_{\mathrm{add}}[X_j\mid X_{\{-j\}}] =
\EE_{\mathrm{add}}[X_j\mid X_{A_j}].
\]
We assume the following:
\begin{longlist}[(B1)]
\item[(B1)] For all $j=1,\ldots,p$: for all $k \in\pa(j)$,
\[
\EE_{\mathrm{add}}\bigl[\bigl(X_j - \EE_{\mathrm{add}}
\bigl[X_j\mid X_{A_j \setminus
k}\bigr]\bigr)\mid  X_{k}\bigr] \not\equiv0.
\]
\end{longlist}
Assumption~\textup{(B1)} requires that for each $j=1,\ldots,p$: $X_k$ [$k \in
\pa(j)$] has an additive influence on $X_j$ given all additive effects from
$X_{A_j \setminus k}$.

%& & X_{j^*} \mbox{has an additive influence on $X_j$ given}\nonumber\\
%& &\mbox{given all additive effects from $X_{C\setminus j^*}$}\nonumber
%& &\forall j^* \in\pa(j), \forall C \subseteq\{1,\ldots, p\}
%j.
%The wording ``additive influence'' can be formalized. Denote the best
%additive function by
%& &\EE_{\mathrm{add}}[X_j\right\vert X_{C\setminus j^*}] = \sum_{k \in
%C\setminus
%j^*}
%g_{jk}^*(X_k),\\
%& &\{g_{jk}^*\}_{k \in C\setminus j^*} = \mbox{argmin}_{\{g_{jk}\}_k}
%Then, $X_{j^*} \mbox{has an additive influence on $X_j$ given given
%all
% additive effects from} X_{C\setminus j^*}$ if and only if
% j^*}])\left\vert X_{j^*}] \not\equiv0.

%le4 #&#
%
\begin{lemm}\label{lemm0}
Assume that \textup{(B1)} holds. Then, for all $j=1,\ldots,p$: $\pa(j)
\subseteq A_j$.
\end{lemm}

A proof is given in the supplemental article \cite{CAMsuppl14}. Lemma
\ref{lemm0} justifies, for the
population case, to pursue preliminary neighborhood selection followed by
restricted maximum likelihood estimation: because
$\pa(j) \subseteq A_j$, the restriction in the maximum likelihood estimator
is appropriate and a true permutation in $\pi^0 \in\Pi^0$ leads to a valid
restriction $R_{\pi^0,j} \supseteq\pa(\pi^0(j))$ (when defined with the
population sets~$A_j$).
%satisfies the restriction $R$ (when defined with the population sets
%$A_j$).(wh%en defined with the population sets $A_j$)
%and a true permutation in $\Pi^0$ satisfies the restriction $R$ (when
%defined with the population sets $A_j$).

For estimation, we assume the following:
\begin{longlist}[(B2)]
\item[(B2)] The selected variables in $\hat{A}_j$ from neighborhood
selection satisfy: with probability tending to~1 as $n \to\infty$,
\begin{enumerate}[(ii)]
\item[(i)] $\hat{A}_j \supseteq A_j$ $(j=1,\ldots,p)$,
\item[(ii)] $\max_{j=1,\ldots,p} \llvert \hat{A}_j\rrvert \le M <
\infty$ for some
positive constant $M < \infty$.
\end{enumerate}
\end{longlist}
Assumption~\textup{(B2)(i)} is a rather standard screening assumption. It holds for
the Group Lasso with sparsity-smoothness
penalty: using a basis expansion as in (\ref{basis-exp}), the
condition is
implied by a sparsity assumption, a group compatibility condition (for
the basis
vectors), and a beta-min
condition about the minimal size of the $\ell_2$-norm of the coefficients
for the basis functions of the active variables in $A_j$; see
\cite{pbvdg11}, Chapter~5.6, Theorem 8.2.
% ich nicht machen.}
The sparsity and group compatibility condition ensure identifiability of
the active set, and hence, they
exclude concurvity (or collinearity) among the additive functions in the
structural equation model.
Assumption~\textup{(B2)(ii)} can be ensured by
assuming $\max_j \llvert A_j\rrvert \le M_1 < \infty$ for some
positive constant
$M_1 <
\infty$ and, for example, group restricted eigenvalue
assumptions for the design matrix (with the given basis); see
\cite{zhang2008sparsity,geer11} for the case without groups.

Finally, we need to strengthen assumption \textup{(A2)} and \textup{(A3)}.
\begin{longlist}[(B3)]
\item[(B3)]
\begin{enumerate}[(iii)]
\item[(i)]For $B \subseteq\{1,\ldots,p\} \setminus j$ with $\llvert
B\rrvert \le M$,
with $M$ as in \textup{(B2)}, denote by $h_{j,g}^{B} = (X_j - \sum_{k
\in B} g_k(X_k))^2$. For some $0 < K < \infty$, it holds that
\[
\max_{j=1,\ldots,p} \max_{B \subseteq\{1,\ldots,p\} \setminus j,
\llvert B\rrvert \le
M} \sup
_{g
\in{\mathcal F}^{\oplus\llvert B\rrvert }} \rho_K\bigl(h_{j,g}^{B}
\bigr) \le D_1 < \infty,
\]
where
\[
\rho_K^2\bigl(h_{j,g}^{B}\bigr) =
2 K^2 \EE_{\theta^0} \bigl[\exp\bigl(\bigl\llvert
h_{j,g}^{B}(X)\bigr\rrvert/K\bigr) -1 - \bigl\llvert
h_{j,g}^{B}(X)\bigr\rrvert/K\bigr].
\]
\item[(ii)] For $V = 1/\alpha$,
\[
\max_{j=1,\ldots,p}\Biggl(\sum_{m=1}^{\infty}
\bigl(M_m^2 \PP[X_j \in I_m]
\bigr)^{V/(V+4)}\Biggr)^{(V+4)/8} \le D_2 < \infty.
\]
This assumption is typically weaker than what we require in \textup{(B3)(i)}, when
assuming that the values $M_m$ are reasonable (e.g., bounded).
\item[(iii)]
\begin{eqnarray*}
\max_j \EE\llvert X_j\rrvert
^4 \le D_3 &<& \infty,
\qquad
\max_j \sup_{f \in{\mathcal F}} \EE\bigl\llvert
f(X_j)\bigr\rrvert^{4} \le D_4 < \infty.
\end{eqnarray*}
\end{enumerate}
\item[(B4)] The error variances satisfy $\min_\pi\min_j
(\sigma_{j}^{\pi,0})^2 \ge L > 0$.
%Consider the basis functions $b_r(\cdot)$ appearing in
%${\mathcal F}_n$\dvtx  for the true functions $f_{j,k}^0 \in{\mathcal F}$,
%we assume an
%expansion
%f_{j,k}^0(x) = \sum_{r=1}^{\infty} a_{f_{j,k}^0;r} b_{j,k;r}(x)
%with smoothness condition\dvtx
%k^{-\beta}.
%and $\max_{j,k} \EE\left\vert b_{j,k;r}(X_k)\left\vert^4 \le M <
%%Note that these conditions ensure that for some $0 < D \le\infty$\dvtx
%%\begin{eqnarray*}
%%\max_j \EE[\left\vert\sum_{k \in\pa_{D^0}(j)} f^{0}_{j,k}(X_k)\left
%f^{0}_{n;j,k}(X_k)\left\vert^{4}] \le D.
%%\end{eqnarray*}
\end{longlist}
Assumption~\textup{(B3)(i)} requires exponential moments. We note that the sum of additive
functions over the set $B$ is finite. Thus, we essentially require
exponential moments for the square of finite sums of additive functions.

%th3 #&#
%
\begin{theo}\label{th3}
Consider an additive structural equation model as in (\ref{modsem})
with independent potentially non-Gaussian errors $\varepsilon_1,\ldots
,\varepsilon_p$ having
$\EE[\varepsilon_j] = 0$ $(j=1,\ldots,p)$. Assume either of the following:
\begin{longlist}[2.]
\item[1.] \textup{(A1)}, \textup{(A4)} and \textup{(B1)--(B4)} hold, and for $\xi_p$ in (\ref{identif2})
(see also Remark \ref{remark2}):
\[
\max \Bigl(\sqrt{\log(p)/n}, \max_{j,k} \EE\bigl[\bigl(f_{j,k}^0(X_k) - f_{n;j,k}^0(X_k)\bigr)^2\bigr]\Bigr) = o(\xi_p).
\]

\item[2.] \textup{(A1)}, \textup{(A4)} and \textup{(B1)--(B4)} hold, and for $\xi_p^{a_n}$ in (\ref{identif3}):
\[
\max \Bigl(\sqrt{\log(p)/n}, \max_{j,k} \EE\bigl[\bigl(f_{j,k}^0(X_k) - f_{n;j,k}^0(X_k)\bigr)^2\bigr]\Bigr) = o\bigl(\xi_p^{a_n}\bigr).
\]
%
%{identif2}) (see also Remark~\ref{remark2}):
%%
%- f_{n;j,k}^0(X_k)\bigr)^2\bigr]
%%
%{identif3}):
%%
%- f_{n;j,k}^0(X_k)\bigr)^2\bigr]
%
%$\liminf_{n \to\infty}
% \xi_{p}^{a_n} \gg???$ in
% formula \eqref{identif3}.
\end{longlist}
Then, for the restricted maximum likelihood estimator in (\ref{rmlepi}):
\[
\PP\bigl[\hat{\pi} \in\Pi^0\bigr] \to1\qquad (n \to\infty).
\]
\end{theo}

A\vspace*{1pt} proof is given in the supplemental article \cite{CAMsuppl14}. The
assumption that $\EE[(f_{j,k}^0(X_k)
- f_{n;j,k}^0(X_k))^2]$ is of\vspace*{1pt} sufficiently small order can be ensured
by the
following condition.
\begin{longlist}[(B$_{\mathrm{add}}$)]
\item[(B$_{\mathrm{add}}$)]
Consider the basis functions $b_r(\cdot)$ appearing in
${\mathcal F}_n$: for the true functions $f_{j,k}^0 \in{\mathcal F}$,
we assume an
expansion
\[
f_{j,k}^0(x) = \sum_{r=1}^{\infty} \alpha_{f_{j,k}^0;r} b_{r}(x)
\]
with smoothness condition:
\[
\sum_{r=k}^{\infty} \bigl\llvert
\alpha_{f_{j,k}^0;r}^0\bigr\rrvert\le C k^{-\beta}.
\]
\end{longlist}
Assuming (B$_{\mathrm{add}}$), we have that $\EE[(f_{j,k}^0(X_k)
- f_{n;j,k}^0(X_k))^2] = O(a_n^{-(\beta- 1 -
\kappa)})$ for any $\kappa> 0$: for example, when using $a_n \to
\infty$ and for
$\beta> 1$, $\EE[(f_{j,k}^0(X_k)
- f_{n;j,k}^0(X_k))^2] \to0$.\vspace*{3pt}
%and $\max_{j,k} \EE\left\vert b_{j,k;r}(X_k)\right\vert^4 \le M <
%%Note that these conditions ensure that for some $0 < D \le\infty$:
%%\begin{eqnarray*}
%%\max_j \EE[\left\vert\sum_{k \in\pa_{D^0}(j)} f^{0}_{j,k}(X_k)\left
%f^{0}_{n;j,k}(X_k)\left\vert^{4}] \le D.
%%\end{eqnarray*}

%We can take a weaker requirement than for
%$\xi_p$ in \eqref{identif2}. Consider
%R_{\mathrm{pop},j} = \{\pi; \pa_{D_{\mathrm{full}}^{\pi}\left\vert
%_{A}}(j)
%where $A = \cup_{j=1}^p A_j$. Denote by
%R_{\mathrm{pop}} = \cap_{j=1}^p R_{\mathrm{pop},j}.
%Then define
% (\Pi^0)^c}p^{-1} \sum_{j=1}^p (\log(\sigma^{\pi,0}_{\pi^{-1}(j)}) -
%and require the bound for $\xi_{\mathrm{R_{\mathrm{pop}}},p}$ instead
%of
%for $\xi_p$.
%And likewise, we could consider a restricted version of $\xi_{p}^{a_n}$
%instead of the latter in \eqref{identif3}.

Uniform convergence can be obtained exactly as described after the
discussion of Theorem~\ref{th1}: when requiring the additional uniform
versions \textup{(U3)--(U4)} [since \textup{(B3)} and \textup{(B4)} invoke already uniform bounds
we do
not need \textup{(U1)} and \textup{(U2)}], and requiring uniform convergence of the
probability in \textup{(B2)}, we obtain uniform convergence over the
corresponding class of distributions analogously as in (\ref{uniform}).

%s5 #&#
\section{Computation and implementation}\label{seccomputation}
In Section~\ref{secmle}, we have decomposed the problem of learning
DAGs from observational data into two main parts: finding the correct
order (Section~\ref{subsecmle}) and feature selection (Section~\ref
{subsecvarsel}).
Our algorithm and implementation consists of two corresponding parts:
\textit{IncEdge} is a greedy procedure providing an
estimate $\hat\pi$ for equation~(\ref{mlepi}) and
\textit{Prune} performs the feature selection.
Section~\ref{secrestrMLE} discusses the benefits of performing a
preliminary neighborhood selection before estimating the causal order,
and we
call the corresponding part \textit{PNS}. The combination \textit{PNS}${}+{}$\textit{IncEdge} provides an estimate for equation (\ref{rmlepi}).

%f1 #&#
%
\begin{figure}%[b]

\includegraphics{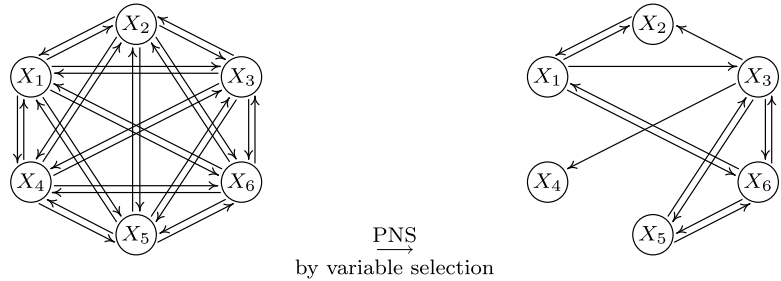}

\caption{Step~\textit{PNS}. For each variable the set of possible parents is
reduced (in this plot, a directed edge from $X_k$ to $X_j$ indicates that
$X_k$ is a selected variable in $\hat{A}_j$ and a possible parent of
$X_j$). This reduction leads to a considerable computational gain in the
remaining steps of the procedure.}
\label{figestcig}
\end{figure}

%f2 #&#
%
\begin{figure}[b]

\includegraphics{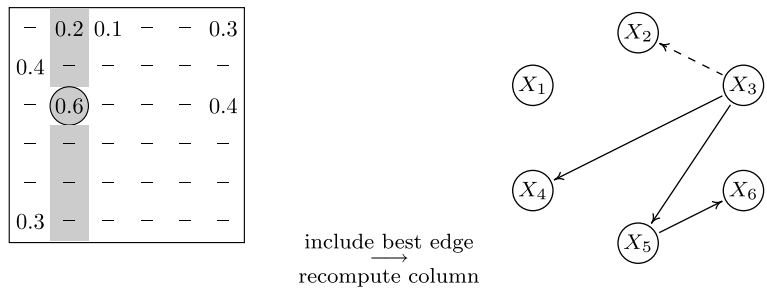}

%shorten >=1pt, shorten <=1pt]
%};
%$\quad\underset{\text{\scriptsize{recompute column}}}{\overset{\text{
%shorten >=1pt, shorten <=1pt]
% \draw(1,2) node(x1) [circle, draw] {$X_1$};
% \draw(2,2.5) node(x2) [circle, draw] {$X_2$};
% \draw(3,2) node(x3) [circle, draw] {$X_3$};
% \draw(1,1) node(x4) [circle, draw] {$X_4$};
% \draw(2,0.5) node(x5) [circle, draw] {$X_5$};
% \draw(3,1) node(x6) [circle, draw] {$X_6$};
% \draw[arcsq-, dashed] (x2) -- (x3);
% \draw[-arcsq] (x3) -- (x4);
% \draw[-arcsq] (x3) -- (x5);
% \draw[-arcsq] (x5) -- (x6);
\caption{Step~\textit{IncEdge}. At each iteration the edge leading to the
largest decrease of the negative log-likelihood is included.}
\label{figincedge}
\end{figure}

%
%f3 #&#
%
\begin{figure}%[b]

\includegraphics{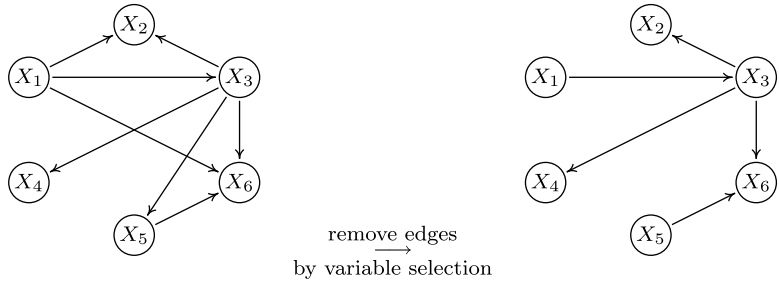}

%shorten >=1pt, shorten <=1pt]
% \draw(1,2) node(x1) [circle, draw] {$X_1$};
% \draw(2,2.5) node(x2) [circle, draw] {$X_2$};
% \draw(3,2) node(x3) [circle, draw] {$X_3$};
% \draw(1,1) node(x4) [circle, draw] {$X_4$};
% \draw(2,0.5) node(x5) [circle, draw] {$X_5$};
% \draw(3,1) node(x6) [circle, draw] {$X_6$};
% \draw[arcsq-] (x2) -- (x3);
% \draw[-arcsq] (x1) -- (x2);
% \draw[-arcsq] (x1) -- (x3);
% \draw[-arcsq] (x1) -- (x6);
% \draw[-arcsq] (x3) -- (x6);
% \draw[-arcsq] (x3) -- (x4);
% \draw[-arcsq] (x3) -- (x5);
% \draw[-arcsq] (x5) -- (x6);
%$\quad\underset{\text{\scriptsize{by variable selection}}}{\overset{
%shorten >=1pt, shorten <=1pt]
% \draw(1,2) node(x1) [circle, draw] {$X_1$};
% \draw(2,2.5) node(x2) [circle, draw] {$X_2$};
% \draw(3,2) node(x3) [circle, draw] {$X_3$};
% \draw(1,1) node(x4) [circle, draw] {$X_4$};
% \draw(2,0.5) node(x5) [circle, draw] {$X_5$};
% \draw(3,1) node(x6) [circle, draw] {$X_6$};
% \draw[arcsq-] (x2) -- (x3);
%% \draw[-arcsq] (x1) -- (x2);
% \draw[-arcsq] (x1) -- (x3);
%% \draw[-arcsq] (x1) -- (x6);
% \draw[-arcsq] (x3) -- (x6);
% \draw[-arcsq] (x3) -- (x4);
%% \draw[-arcsq] (x3) -- (x5);
% \draw[-arcsq] (x5) -- (x6);
\caption{Step~\textit{Prune}. For each node, variable selection
techniques are exploited to remove nonrelevant edges.}
\label{figprune}
\end{figure}

The three components of our implementation are described in the
following subsections, Figures~\ref{figestcig},~\ref{figincedge}
and~\ref{figprune} present the steps graphically.
We regard the modular structure of the implementation as an advantage;
each of
the three parts could be replaced by an alternative method (as
indicated in
the subsections below).

%s5.1 #&#
\subsection{Preliminary neighborhood selection: \textit{PNS}}\label{secestcig}
As described in Sec-\break tion~\ref{secrestrMLE}, we fit an additive model for
each variable $X_j$ against all other variables $X_{\{-j\}}$. We
implement this
with a boosting method for additive model fitting
\cite{pbyu03,BuhlmannHothorn06}, using the R-function \texttt{gamboost}
from the package \texttt{mboost}
\cite{hot10}. We select the ten variables that have been
picked most often during $100$ iterations of the boosting method;
hereby, we only
consider variables that have been picked at
least three times during the iterations.
The sets $\hat A_j$ obtained by this
procedure estimate
$A_j \supseteq\pa(j)$ as shown in Lemma~\ref{lemm0}.
%the Markov blanket of node $X_j$ in the underlying DAG.
We construct a graph in which for each $j$, the set $\hat A_j$ is the
parental set for node $j$ corresponding to the variable $X_j$.
Figure~\ref{figestcig} summarizes this step.
We say that the set of ``possible parents'' of node $j$ has been
reduced to the set $\hat A_j$.
%conditional independence graph (CIG).
Importantly, we do not
disregard true parents if the sample size is large enough
(Section~\ref{subsecrmletheory}, Lemma~\ref{lemm0}).
Instead of the boosting method, we could alternatively use additive model
fitting with a sparsity- or sparsity-smoothness penalty
\cite{ravik09,meieretal09}.

% important? Maybe delete and just show undirected skeleton in the left
%and
%right panel}
%gezeichnet!}

%s5.2 #&#
\subsection{Estimating the correct order by greedy search: \textit{IncEdge}}
Let us first consider the situation without \textit{PNS}.
Searching over all permutations $\pi$ for finding $\hat{\pi}$ in
(\ref{mlepi}) is computationally infeasible if the number of
variables $p$ is large.
We propose a greedy estimation procedure that starts with an empty DAG and
adds at each iteration the edge $k \to j$ between nodes $k$ and $j$ that
corresponds to the largest gain in log-likelihood. We therefore compute the
score function in~(\ref{mlepi}), with $\pi$ corresponding to the
current DAG,
\[
\sum_{j=1}^p \log\bigl(\hat{
\sigma}^{\pi}_j\bigr) = \sum_{j=1}^p
\log\Biggl( \Biggl\llVert X_j^{\pi} - \sum
_{k=1}^{j-1} \hat f^{\pi
}_{j,k}
\bigl(X^{\pi}_k\bigr)\Biggr\rrVert_{(n)} \Biggr)
\]
and construct a matrix,
whose entry $(k,j)$ specifies by how
much this score is reduced after adding the edge $k \rightarrow j$ and,
therefore, allowing a nonconstant function $f_{j,k}$ (see Figure~\ref
{figincedge}).
For implementation, we employ additive model fitting with penalized regression
splines (with ten basis functions per variable), using the R-function
\texttt{gam} from the \texttt{R}-package \texttt{mgcv}, in order to
obtain estimates $\hat{f}_{j,k}$ and $\hat\sigma_j$.
After the addition of an edge, we only need to recompute the $j$th
column of the score matrix (see Figure~\ref{figincedge}) since the
score decomposes over all nodes. In order to avoid cycles, we remove
further entries of the score matrix.
After $p(p-1)/2$ iterations, the graph has been completed to a fully
connected DAG.
The latter corresponds to a unique permutation $\hat\pi$.
This algorithm is computationally rather efficient and can easily handle
graphs of up to $30$ nodes without \textit{PNS} (see Section~\ref{secpns}).

If we have performed \textit{PNS} as in Section~\ref{secestcig} we
sparsify the score matrix from the beginning.
%may be confusing with step 3 of the algorithm}.
We only consider entries $(k,j)$ for which $k$ is considered to be a
possible parent of $j$. This way the algorithm is feasible for up to a
few thousands of nodes (see Section~\ref{seccomp}).

Alternative methods for (low-dimensional) additive model fitting include
backfitting \cite{mammen06}, cf.

%s5.3 #&#
\subsection{Pruning of the DAG by feature selection: \textit{Prune}} \label{secprune}

%Fall wo
% man einen full DAG pruned. Hier ist es ein bisschen anders
%geschrieben;
% uynd wahrscheinlich pruned ihr ja nur den restricted DAG, oder? Sollen
% wir Section 2.5 noch klarer machen?}
%ist Sect%ion 2.5. eine Methode, in die man JEDEN dag reinstecken kann.
%Daher war die Ueb%erlegung, dies $S(\cdot)$ zu nennen. Derzeit heisst
%die Operation $\hat\cdot$,% das geht vermutlich auch. Ich habe
%Section 2.5. mal ganz leicht veraendert.}

Section~\ref{subsecvarsel} describes sparse regression techniques for
pruning the DAG that has been estimated by step~\textit{IncEdge}; see
Figure~\ref{figprune}. We implement this task by applying significance
testing of covariates, based on the R-function \texttt{gam}
from the \texttt{R}-package \texttt{mgcv} and declaring significance
if the
reported $p$-values are lower or equal to $0.001$, independently of the
sample size (for problems with small sample size, the $p$-value threshold
should be increased).
%Many other choices would be possible, too.

If the DAG estimated by (\textit{PNS} and) \textit{IncEdge} is a super DAG of the
true DAG, the estimated
interventional distributions are correct; %\Peter{for the full DAG or
%as long as we do not prune away too many edges},
see
Section~\ref{subseccausalconsist}.
%and therefore the structural intervention distance described in
%Section~
%is zero.
This does not change if \textit{Prune} removes additional ``superfluous'' edges.
The structural Hamming distance to the true graph, however, may reduce
significantly after performing \textit{Prune}; see Section~\ref{secpns}.
Alternative methods for hypothesis testing in (low-dimensional) additive
are possible \cite{wood06}, cf., or one could use penalized
additive model fitting for variable selection
\cite{YI06,ravik09,meieretal09}.

%s6 #&#
\section{Numerical results}\label{secnumerical}

%s6.1 #&#
\subsection{Simulated data}
We show the effectiveness of each step in our algorithm
(Section~\ref{secpns}) and compare the whole procedure to other state
of the art methods (Section~\ref{seccomp}). We investigate
empirically the role of noninjective functions (Section~\ref
{secinj}) and discuss the linear Gaussian case (Section~\ref
{secling}). In Section~\ref{secmis}, we
further check the robustness of our method against model misspecification,
that is, in the case of non-Gaussian noise or nonadditive functions. For
evaluation, we compute the structural intervention distance that we
introduce in Section~\ref{secsid}.

For simulating data, we randomly choose a correct ordering $\pi^0$ and
connect each pair of variables (nodes) with a probability
$p_{\mathrm{conn}}$. If not
stated otherwise, each of the possible $p(p-1)/2$ connections is included
with a probability of $p_{\mathrm{conn}} = 2/(p-1)$ resulting in a
sparse DAG
with an expected number of $p$ edges. Given the structure, we draw the
functions $f_{j,k}$ from a Gaussian process with a Gaussian (or RBF)
kernel with bandwidth one
and add
Gaussian noise with standard deviation uniformly sampled between $1/5$
and $\sqrt{2}/5$. All
nodes without parents have a standard deviation between $1$ and $\sqrt
{2}$. The
experiments are based on~100 repetitions if the description does not
say differently. %\Jan{Passt das hierhin oder soll ich das bei jedem
%Exp. einzeln hinschreiben?}
% \Jonas{This looks a bit funny, shall we repeat the experiments with
%different values?}.

%f4 #&#
%
\begin{figure}[b]

\includegraphics{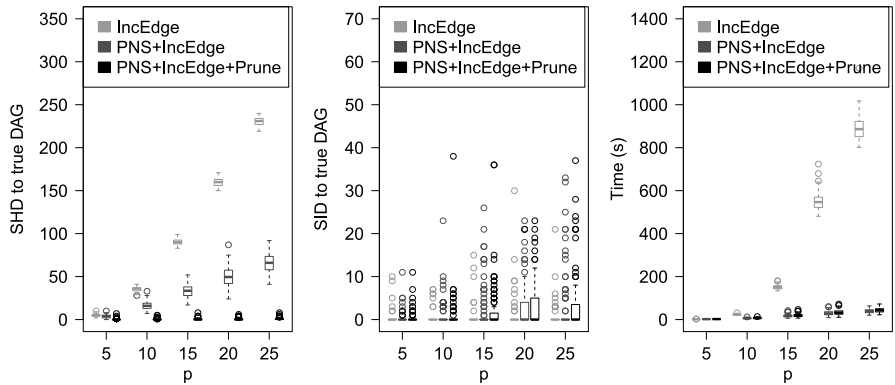}

\caption{The plots show the effect of the individual steps of our
method. \textit{Prune} reduces the SHD to the true DAG but leaves the SID
almost unchanged. \textit{PNS} reduces the computation time, especially for
large $p$.}
\label{figpns}
\end{figure}

All code is provided on the second author's homepage.

%s6.1.1 #&#
\subsubsection{Structural intervention distance} \label{secsid}
As a performance measure, we consider the recently proposed structural
intervention distance (SID); see \cite{peterspbSID13}. The SID is well
suited for quantifying the correctness of an order among variables, mainly
in terms of inferring causal effects afterward.
%$\hat{\pi}$,
%consider the fully connected DAG $D_{\mathrm{full}}^{\hat{\pi}}$ which
%is
%compatible with the ordering and permutation $\hat{\pi}$, i.e.,
%there is an
%edge $\hat{\pi}(j) \to\hat{\pi}(k)$ in
%$\hat{D}_{\mathrm{full}}^{\hat{\pi}}$ if and only if $k < j$.
%%\Jonas{$D_{\mathrm{full}}^{\hat{\pi}}$} if and only if $\hat{\pi}(j) <
%%\hat{\pi}(k)$.
%A correct order $\pi^0 \in\Pi^0$ then satisfies
%$\hat{D}_{\mathrm{full}}^{\pi^0} \supseteq D^0$
%%\Jonas{$D_{\mathrm{full}}^{\pi^0} \supseteq D^0$}\dvtx  the latter is the
%property
%in \eqref{screening} which allows for consistent estimation of causal
%effects as described in Section~\ref{subseccausalconsist}.
%%The SID roughly
%%counts the number of edges where $\hat{D}_{\mathrm{full}}^{\hat{\pi}}
%%\supseteq D^0$ is violated\dvtx  because it only involves the fully
%connected DAG
%%$\hat{D}_{\mathrm{full}}^{\hat{\pi}}$,
%The SID is a measure for assessing the quality of an ordering.
%der Satz% ``The SID is a measure for assessing the quality of an
%ordering.'' stand. Den %wollte ich erlaeutern. Und sagen, warum die
%SID(wahrer DAG, wahre Permutation) %null ist. Kann wegen mir wirklich
%weg oder verkuerzt werden zu etwas wie\dvtx \\
It counts the number of wrongly estimated causal
effects. Thus, the SID between the true DAG $D^0$ and the fully connected
DAGs corresponding to the true permutations $\pi^0 \in\Pi^0$ is
zero; see
Section~\ref{subseccausalconsist}.
%Subsection?}
%Section~\ref{secpns} shows that the SID is
%almost unaffected by the feature selection step with penalized
%regression and hence, regularization for feature selection is only
%needed for
%selecting a right-sized model.
%%\Peter{Kommentar 26. vom referee\dvtx  was machen wir da?}
%%\Jonas{Verstehe ich nicht. Vielleicht ist es nicht Kommentar~26?}

%s6.1.2 #&#
\subsubsection{Effectiveness of preliminary neighborhood selection and pruning} \label{secpns}
We demonstrate the effect of the individual steps of our algorithm.
Figure~\ref{figpns} shows the performance (in terms of SID and SHD)
of our method and the corresponding time consumption (using eight
cores) depending on which of the steps are performed. If only \textit{IncEdge}
is used, the SHD is usually large because the output is a fully
connected graph. Only after the step~\textit{Prune} the SHD becomes
small. As
discussed in Section~\ref{subseccausalconsist} the pruning does not
make a big difference for the SID.
Performing these two steps is not feasible for large $p$. %\Jonas{at
%Jan\dvtx  How many cores??} \Jan{OK}
The time consumption is reduced significantly if we first apply the
preliminary neighborhood selection \textit{PNS}. In particular, this first
step is required in the case of $p>n$ in order to avoid a degeneration
of the score function.

%s6.1.3 #&#
\subsubsection{Comparison to existing methods} \label{seccomp}
\begin{figure}[b]%[b]

\includegraphics{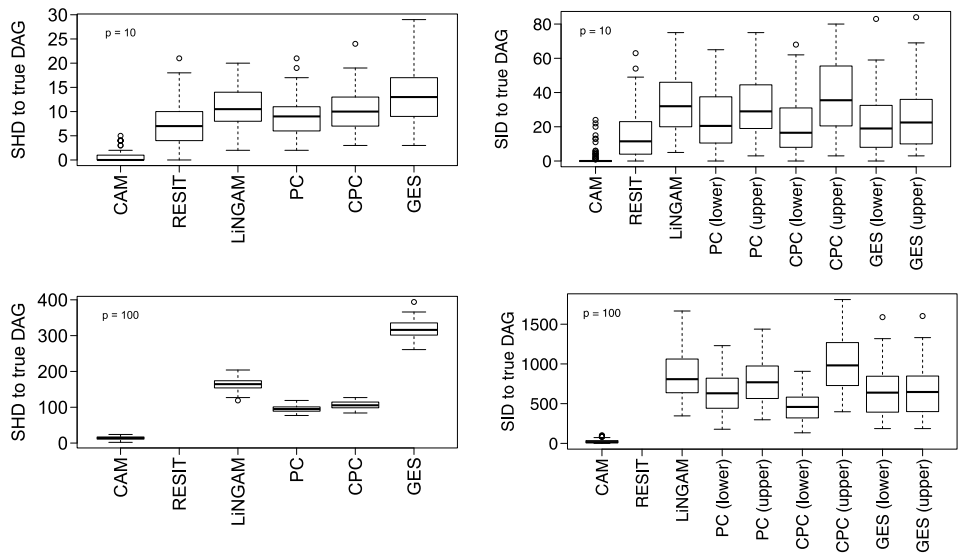}

\caption{SHD (left) and SID (right) for different methods on sparse
DAGs with $p=10$ (top) and $p=100$ (bottom); the sample size is $n=200$.}
\label{figcomp}
\end{figure}

Different procedures have been proposed to address the problem of inferring
causal graphs from a joint observational distribution. We compare the
performance of our method to greedy equivalence search (GES)
\cite{chick02}, the PC algorithm \cite{sgs00}, the conservative PC
algorithm (CPC)~\cite{ram06}, LiNGAM~\cite{shim06} and regression with
subsequent independence tests (RESIT) \cite{moo09,petersetal13}. The
latter has been used with a significance level of $\alpha= 0$, such that
the method does not remain undecided. Both PC methods are equipped with
$\alpha= 0.01$ and partial correlation as independence test. GES is used
with a linear Gaussian score function. Thus, only RESIT is able to model
the class of nonlinear additive functions.
We apply the methods to DAGs of size $p=10$ and $p=100$, whereas in both
cases, the sample size is $n=200$. RESIT is not applicable for graphs with
$p=100$ due to computational reasons. Figure~\ref{figcomp} shows that our
proposed method outperforms
the other approaches both in terms of SID and SHD.

%f5 #&#
%

The difference between the methods becomes even larger for dense graphs
with an expected number of $4p$ edges and strong varying degree of nodes
(results not shown).

Only the PC methods and the proposed method CAM scale to high-dimensional
data with $p=1000$ and $n=200$. Keeping the same (sparse) setting as
above results
in SHDs of $1214 \pm37$, $1330 \pm40$ and $477 \pm19$ for PC, CPC and
CAM, respectively. These results are based on five experiments.

%s6.1.4 #&#
\subsubsection{Injectivity of model functions} \label{secinj}
\begin{figure}[b]

\includegraphics{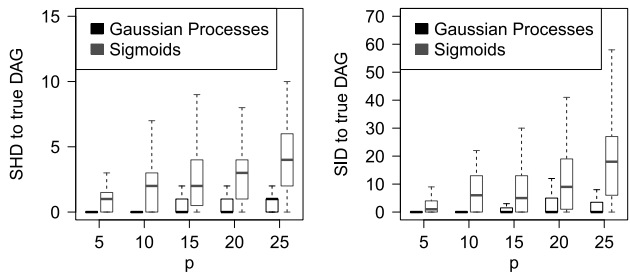}

\caption{SHD (left) and SID (right) for various values of $p$ and
$n=300$. The plots compare the performances of CAM for the additive SEM
(\protect\ref{SEMadd}) with functions generated by Gaussian
processes (noninjective in general) and sigmoid-type functions (injective).}
\label{figinjec}
\end{figure}

In general, the nonlinear functions that are generated by Gaussian
processes are not injective.
We therefore test CAM for the case where every function in (\ref{SEMadd})
is injective. Correct direction of edges $(j,k)$ is a more difficult task
in this setting. We sample
sigmoid-type functions of the form
\[
f_{j,k}(x_k)=a \cdot\frac{b \cdot(x_k+c)}{1+\llvert b \cdot
(x_k+c)\rrvert } %
\]
with $a \sim\operatorname{Exp}(4)+1$, $b \sim\mathcal
{U}([-2,-0.5]\cup[0.5,2])$
and $c \sim\mathcal{U}([-2,2])$; as before, we use Gaussian noise.
Note that
some of these functions may be very close to linear functions which
makes the direction of the corresponding edges difficult to identify.
Figure~\ref{figinjec} shows a
comparison of the performance of CAM in the previously applied setting with
Gaussian processes and in the new setting with sigmoid-type functions. As
expected, the performance of CAM decreases in this more difficult
setting but is still better than for the competitors such as RESIT,
LiNGAM, PC,
CPC and GES (not shown).
%mal getestet?}

%f6 #&#
%

%f7 #&#
%
\begin{figure}%[b]

\includegraphics{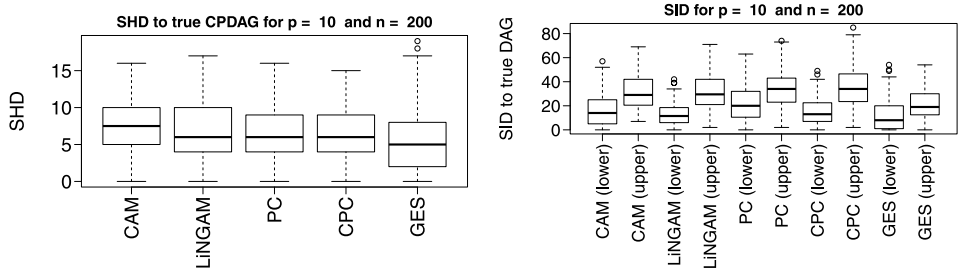}

\caption{Data are generated by linear Gaussian SEM. SHD between true
and estimated CPDAG (left), lower and upper bounds for SID between true
DAG and estimated CPDAG (right).}
\label{figlin}
\end{figure}

%s6.1.5 #&#
\subsubsection{Linear Gaussian SEMs} \label{secling}
In the linear Gaussian setting, we can only identify the Markov
equivalence class of the true graph (if we assume faithfulness). We now
sample data from a linear Gaussian SEM and expand the DAGs that are
estimated by CAM and LiNGAM to CPDAGs, that is, we consider the
corresponding Markov equivalence classes. The two plots in Figure~\ref
{figlin} compare the different methods for $p=10$ variables and
$n=200$. They show the structural Hamming distance (SHD) between the
estimated and the true Markov equivalence class (left), as well as
lower and upper bounds for the SID (right). (By the definition of lower
and upper bounds of the SID, the SID between the true and estimated DAG
lies in between those values.) The proposed method has a disadvantage
because it uses nonlinear regression instead of linear regression. The
performance is nevertheless comparable. Remark~\ref{rem1} discusses
that at least in principle, this scenario is
detectable.

%
%Oder wie kommt man sonst zu den SID bounds?}

%s6.1.6 #&#
\subsubsection{Robustness against nonadditive functions and non-Gaussian errors} \label{secmis}
This work focuses on the additive model (\ref{SEMadd}) and Gaussian noise.
The score functions~(\ref{mlepi})
and~(\ref{rmlepi}) and their corresponding optimization problems
depend on these model assumptions.
The DAG remains identifiable (under weak assumptions) even if the functions
of the data generating process are not additive or the noise variables
are non-Gaussian \cite{petersetal13}, cf.
The following experiments analyze the empirical performance of our
method under these misspecifications. The case of misspecified error
distributions is discussed in Section~\ref{secmissp}.

\begin{figure}[t]

\includegraphics{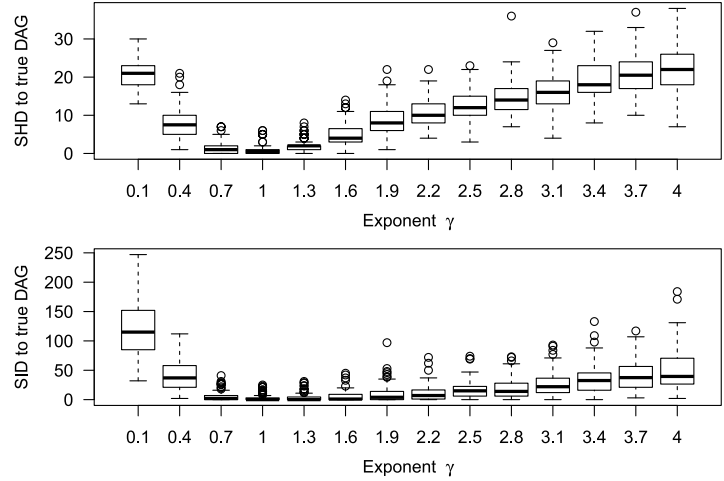}

\caption{SHD (top) and SID (bottom) for $p=25$ and $n=300$ in the case
of misspecified
models. The plot shows deviations of the noise from a normal
distribution (only $\gamma=1$ corresponds to Gaussian noise).
%The lower panel shows the SID for different numbers of basis functions
%used to fit the gam models
}
\label{figrob}
\end{figure}

As a first experiment, we examine
deviations from the Gaussian noise assumption by setting $\varepsilon
_j =
\operatorname{sign}(N_j) \llvert N_j\rrvert ^{\gamma} $ with $N_j \sim
{\mathcal
N}(0,\sigma_j^2)$
for different\vspace*{2pt} exponents $0.1 \leq\gamma\leq4$. Only $\gamma= 1$
corresponds to normally distributed noise. Figure~\ref{figrob} shows
the change in SHD and SID when varying $\gamma$.
%As a second experiment we fix the number of basis functions used in the
%steps {\it IncEdge} and {\it Prune} of our algorithm. The recorded SIDs
%are shown in Figure~\ref{figrob} (lower). \Jan{Irgendwie ist es
%schwierig, aus diesem Experiment einen geeigneten Schluss zu ziehen...}
%f8 #&#
%

%f9 #&#
%
\begin{figure}[b]

\includegraphics{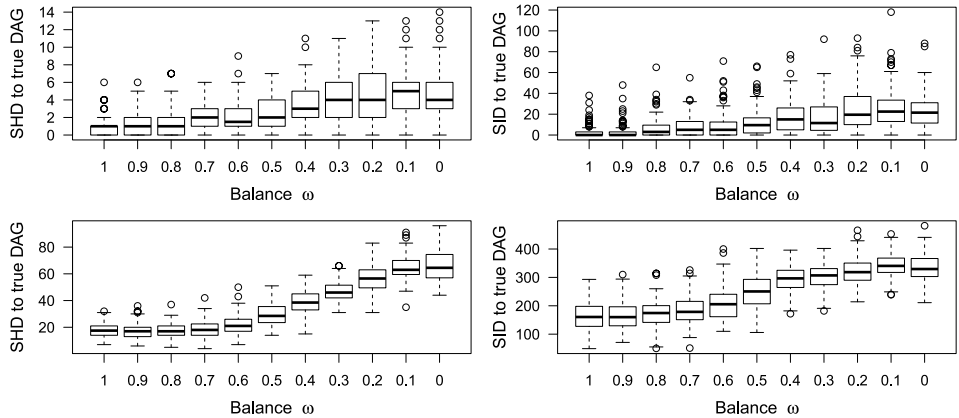}

\caption{SHD (left) and SID (right) for $p=25$ and $n=300$ in the case
of misspecified
models. The plot shows deviations from additivity for sparse (top) and
nonsparse (bottom) truths, respectively (only $\omega=1$ corresponds
to a fully additive model).}
\label{figrob2}
\end{figure}

%f10 #&#
%
\begin{figure}[b]

\includegraphics{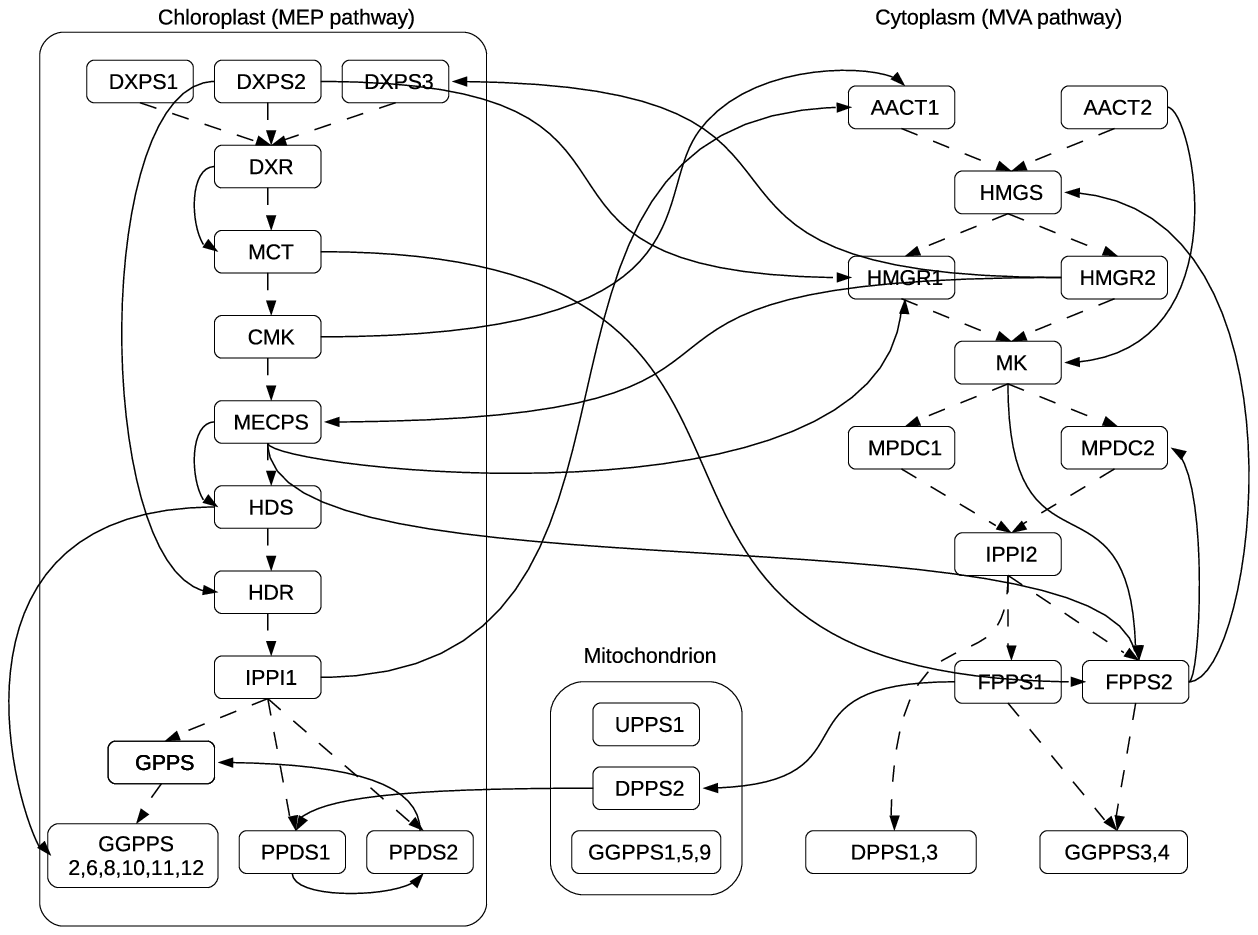}

\caption{Gene expressions in isoprenoid pathways. The twenty best scoring
edges provided by the method CAM.}\label{figtop20}
\end{figure}

As a second experiment, we examine deviations from additivity by simulating
from the model
\[
X_j = \omega\cdot%\left(
\sum_{k \in\pa_{D}(j)}
f_{j,k}(X_k) %\right)
+ (1-\omega) \cdot
f_j(X_{\pa_{D}(j)}) + \varepsilon_j
\]
for different values of $\omega\in[0,1]$ and Gaussian noise.
Both, $f_{j,k}$ and $f_j$ are drawn from a Gaussian process using an
RBF kernel with bandwidth one.
Note that $\omega=1$ corresponds to the fully additive model~(\ref
{modsem}), whereas for $\omega=0$, the value of $X_j$ is given as a
nonadditive function of all its parents. Figure~\ref{figrob2} shows
the result for a sparse truth with expected number of $p$ edges (top)
and a nonsparse truth with expected number of $4p$ edges (lower). In
sparse DAGs, many nodes have a small number of parents and our
algorithm yields a comparably small SID even if the model
%misspecified and each node is given as a
contains nonadditive functions.
% of all its parents.
If the underlying truth is nonsparse, the performance of our algorithm
becomes worse but it is still slightly better than PC which achieves average
lower bounds of SID values of roughly $520$, both for $\omega= 1$ and for
$\omega= 0$ (not shown).
%Figure~\ref{figrob} (left) shows the drop in performance when the
%noise

%s6.2 #&#
\subsection{Real data}

\begin{figure}

\includegraphics{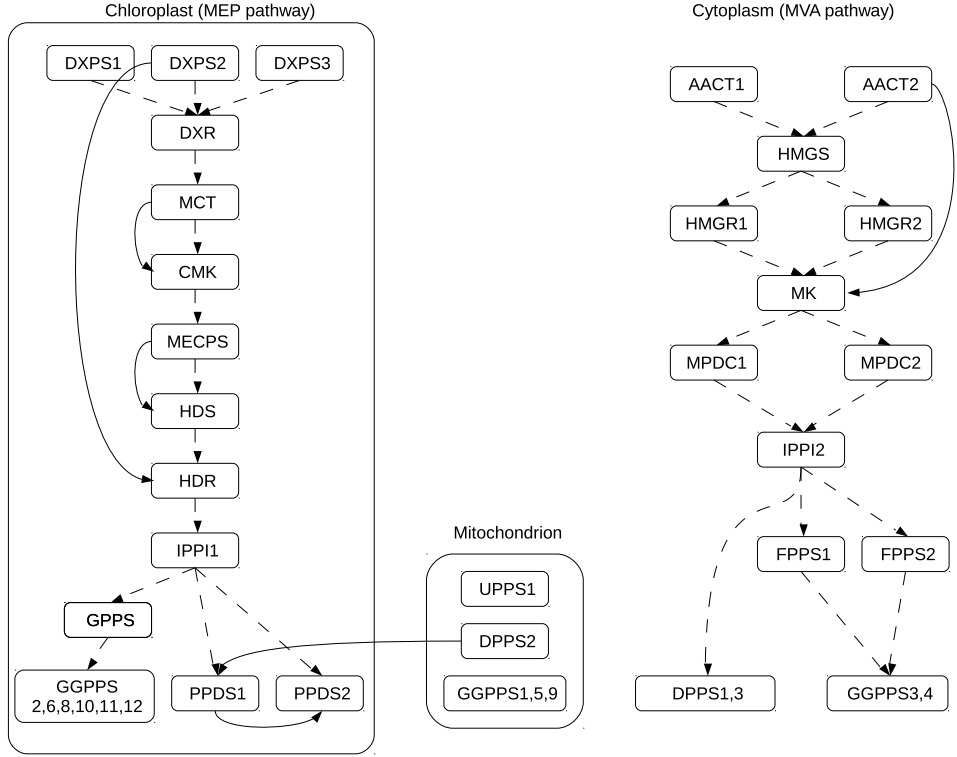}

\caption{Gene expressions in isoprenoid pathways. Edges estimated by
stability selection: all directions are in correspondence with the
direction of the pathways.}
\label{figstabsel}
\end{figure}

We apply our methodology to microarray data described in~\cite{wil04}. The
authors concentrate on $39$ genes ($118$ observed samples) on two
isoprenoid pathways in Arabidopsis thaliana.
The dashed edges in Figures~\ref{figtop20} and~\ref{figstabsel}
indicate the causal direction within each pathway.
%We reuse both their assignment
%of the genes to the pathways and their ordering of the genes within the
%pathway.
While graphical Gaussian models are applied to
estimate the underlying interaction network by an undirected model in
\cite{wil04}, our CAM procedure estimates the structure by a directed
acyclic graph.

%f11 #&#
%

Given a graph structure, we can compute $p$-value scores as described in
Section~\ref{secprune}. Figure~\ref{figtop20} shows the twenty best
scoring edges of the graph estimated by our proposed method CAM (the scores
should not be interpreted as $p$-values anymore since the graph has been estimated
from data).
We also apply stability selection \cite{mein10} to this data set. We
therefore consider $100$ different subsamples of size $59$ and record the
edges that have been considered at least $57$ times as being among the $20$
best scoring edges. Under suitable assumptions, this leads to an
expected number of false positives being less than two
\cite{mein10}. These edges are shown in Figure~\ref{figstabsel}.
They
connect genes within one of the two pathways and their directions agree
with the overall direction of the pathways. Our findings are therefore
consistent with the prior knowledge available. The link MCT${}\rightarrow{}$CMK does not appear in Figure~\ref{figtop20} since it was
ranked as the $22$nd best scoring edge.

%s7 #&#
\section{Conclusions and extensions}\label{secconclusion}

We have proposed maximum likelihood estimation and its restricted version
for the class of additive structural equation models (i.e., causal additive
models, CAMs) with Gaussian errors where the causal structure (underlying
DAG) is identifiable from the observational probability distribution
\cite{petersetal13}. A key component of our approach is to
decouple order search among the variables from feature or edge
selection in
DAGs. Regularization is only necessary for the latter while estimation of
an order can be done with a nonregularized (restricted) maximum likelihood
principle. Thus, we have substantially simplified the problem of structure
search and estimation for an important class of causal models. We established
consistency of the (restricted) maximum likelihood estimator for low- and
high-dimensional scenarios, and we also allow for misspecification of the
error distribution. Furthermore, we developed an efficient computational
algorithm which can deal with many variables, and the new method's accuracy
and performance is illustrated with a variety of empirical results for
simulated and real data. We found that we can do much more accurate
estimation for identifiable, nonlinear CAMs than for nonidentifiable
linear Gaussian structural equation models.

%s7.1 #&#
\subsection{Extensions}\label{subsecextensions}

The estimation principle of first pursuing order search based on
nonregularized maximum likelihood and then using
penalized regression for feature selection works with other structural
equation models where the underlying DAG is identifiable from the
observational distribution. Closely related examples include nonlinearly
transformed additive structural equation models \cite{zhang09} or Gaussian
structural equation models with same error variances \cite{petbu13}.

If the DAG $D$ is nonidentifiable from the distribution $P$, the
methodology needs to be adapted; see\vspace*{1pt} also Remark~\ref{rem1} considering
the linear Gaussian SEM. The true orders $\Pi^0$ can be defined as the
set of
permutations which lead to most sparse autoregressive representations
as in
(\ref{triangular}): assuming faithfulness, these orders correspond to the
Markov equivalence class of the underlying DAG. Therefore, for estimation,
we should use regularized maximum likelihood estimation leading to sparse
solutions with, for example, the $\ell_0$-penalty \cite{chick02,sarpet12}.

Finally, it would be very interesting to extend (sparse) permutation
search to
(possibly nonidentifiable) models with hidden variables
\cite{sgs00,pearl00,Janzing2009,colombetal12} or with graph structures
allowing for
cycles \cite{spirtes95,richardson96,moojietal11,MooijHeskesUAI13}.
Note that unlike linear
Gaussian models, CAMs
are not closed under marginalization: if $X, Y$ and $Z$ follow a
CAM~(\ref{SEMadd}),
then $X$ and $Y$ do not necessarily remain in the class of CAMs.

\section*{Acknowledgments}
The authors thank Richard Samworth for fruitful discussions regarding the
issue of closedness of subspaces allowing to construct proper projections.

\begin{supplement}[id=suppA]
\stitle{Supplement to ``CAM: Causal additive models, high-dimensional order search and penalized regression''}
\slink[doi]{10.1214/14-AOS1260SUPP} %[doi,text={...}] - jei reikia
%suskaldyti doi
\sdatatype{.pdf}
\sfilename{aos1260\_supp.pdf}
\sdescription{This supplemental article \cite{CAMsuppl14} contains all proofs.}
\end{supplement}

% imsref loaded by linak, 2014-09-08 10:27:32
%

\printaddresses
\end{document}